\documentclass[11pt,a4paper]{article}

\usepackage{jheppub}

\usepackage{slashed}
\usepackage{diagbox}
\usepackage{subfigure}

\usepackage{wick}

\DeclareMathOperator{\Tr}{Tr}
\def\II{\hbox{{1}\kern-.25em\hbox{l}}}

\newcommand \vev [1] {\langle{#1}\rangle}

\newcommand{\derright}{\stackrel{\rightarrow}{D}}

\newcommand{\LO}{{\rm\scriptscriptstyle LO}}
\newcommand{\NLO}{{\rm\scriptscriptstyle NLO}}
\newcommand{\WW}{{\rm\scriptscriptstyle WW}}

\newcommand{\Figa}{{\rm\scriptscriptstyle Fig 1a}}

\newcommand{\mI}{{\rm\scriptscriptstyle I}}
\newcommand{\mII}{{\rm\scriptscriptstyle I\!I}}
\newcommand{\mIII}{{\rm\scriptscriptstyle I\!I\!I}}

\def \e {\mbox{e}}

\allowdisplaybreaks

\newcounter{MBQ}

\title{\boldmath 
Radiative leptonic decay $B\to \gamma \ell \nu_\ell$ with subleading 
power corrections}

\author[a]{M. Beneke,} 

\author[b]{V. M. Braun,}

\author[b]{Yao Ji}

\author[a,c]{and Yan-Bing Wei}

\subheader{
\begin{flushright}
{\small
TUM-HEP-1135/18\\
April 12, 2018
}
\end{flushright}
}

\affiliation[a]{
   Physik Department T31, James-Franck-Stra\ss e 1, 
   Technische Universit{\"a}t M{\"u}nchen,\\
   D-85748 Garching, Germany}

\affiliation[b]{
   Institut f\"ur Theoretische Physik, Universit\"at
   Regensburg, \\ D-93040 Regensburg, Germany}

\affiliation[c]{Institute of High Energy Physics, CAS, 
P.O. Box 918(4) Beijing 100049, China}

\emailAdd{vladimir.braun@ur.de}

\emailAdd{yao.ji@ur.de}

\emailAdd{weiyb@ihep.ac.cn}

\abstract{
We reconsider the QCD predictions for the radiative decay 
$B\to \gamma \ell \nu_\ell$ with an energetic photon in the final state  
by taking into account the $1/E_\gamma, 1/m_b$ power-suppressed 
hard-collinear and soft corrections from higher-twist $B$-meson 
light-cone distribution amplitudes (LCDAs). The soft contribution is 
estimated through a dispersion relation and light-cone QCD sum 
rules. The analysis of theoretical uncertainties and the dependence 
of the decay form factors on the leading-twist LCDA $\phi_+(\omega)$ shows 
that the latter dominates. The radiative leptonic decay is therefore well 
suited to constrain the parameters of $\phi_+(\omega)$, including 
the first inverse moment, $1/\lambda_B$, from the expected high-statistics 
data of the BELLE II experiment.
}

\keywords{B physics, NLO computations, nonperturbative effects}

\notoc
\begin{document}

\maketitle

\newpage

%
\section{Introduction}
%

When the energy $E_\gamma$ of the photon is large compared to the 
strong interaction scale $\Lambda$, 
the radiative leptonic decay $B\to \gamma \ell\nu_\ell$ of the charged 
$B$ meson is the simplest decay that probes the light-cone structure 
of the $B$ meson, relevant to QCD factorization of exclusive 
$B$ decays \cite{Beneke:1999br}. In this respect, this decay represents the 
analogue of the $\gamma\gamma^*\to \pi^0$ form factor for mesons with 
a heavy quark, in which case the mass $m_b$ of the quark sets the scale of 
the hard interaction. Factorization at leading power in an expansion 
of the decay amplitude in $\Lambda/E_\gamma$ and $\Lambda/m_b$ has been 
established \cite{Lunghi:2002ju,Bosch:2003fc} to all orders in the 
strong coupling $\alpha_s$. In this approximation, the branching fraction 
depends only on the leading-twist $B$-meson light-cone distribution amplitude 
(LCDA) $\phi_+(\omega)$ \cite{Grozin:1996pq,Beneke:2000wa}. More precisely, 
it is proportional to the square of the inverse moment $1/\lambda_B^2$, 
which is the most important $B$-meson LCDA parameter in exclusive decays. 
Yet, $\lambda_B$ remains uncertain by a large factor with estimates 
ranging from 200~MeV favoured by non-leptonic 
decays \cite{Beneke:2003zv,Beneke:2009ek} to 
$460\pm 110$~MeV from QCD sum rules \cite{Braun:2003wx}. The radiative 
leptonic decay has therefore been suggested as a measurement of 
$\lambda_B$~\cite{Beneke:2011nf}. Including 
next-to-leading logarithmic resummed radiative corrections, known 
next-to-leading power effects and an estimate of an unknown 
next-to-leading power form factor $\xi(E_\gamma)$, the 
partial branching fractions $\mbox{Br}(B\to\gamma\ell\nu_\ell, 
E_\gamma>E_{\rm cut})$ have been predicted in \cite{Beneke:2011nf} 
and have been employed by the BELLE collaboration to provide 
a constraint on $\lambda_B$ from their complete data 
set \cite{Heller:2015vvm}. The main limitation of this method is 
due to $\Lambda/E_\gamma$ and $\Lambda/m_b$ power corrections.

In this paper we attempt to quantify the leading power-suppressed effects. 
A factorization analysis of the  radiative leptonic decay 
in next-to-leading power would be desirable 
and interesting by itself, but 
this can presently not be done with rigour comparable to leading power 
due to a lack of understanding of ``endpoint contributions'' in the 
LCDAs, where the spectator partons in the $B$ meson carry an anomalously 
small momentum fraction $\omega \ll \Lambda$. We therefore resort 
to the light-cone sum rule technique~\cite{Balitsky:1989ry}, 
which expresses the contribution of the 
endpoint region in the partonic calculation through a dispersion 
relation in terms of hadronic resonance parameters and $B$-meson 
LCDAs. This technique was originally applied to the analogous problem 
for the $\gamma\gamma^*\to \pi^0$ form factor 
\cite{Khodjamirian:1997tk, Agaev:2010aq} and for the problem at hand 
in \cite{Braun:2012kp} in the tree-level and leading-twist approximation 
for the $B$-meson LCDAs. The one-loop correction to the 
leading-twist approximation for the dispersive representation of the 
soft contribution was added in \cite{Wang:2016qii}. The 
reanalysis~\cite{Wang:2016qii}
of the predicted branching fraction including these new contributions 
led to a considerable weakening of the bounds on $\lambda_B$. The purpose 
of the present paper is twofold: first, we focus on power corrections 
from higher-twist $B$-meson LCDAs using the complete parametrization 
of these LCDAs from \cite{Braun:2017liq}. Second, we perform an extensive 
analysis of the model dependence by quantifying the uncertainty through 
different families of $B$-meson LCDA models with a consistent implementation 
of the equation-of-motion constraints. Taken together, this results in 
a more reliable assessment of the potential of radiative leptonic decay 
for determining the inverse-moment parameter $\lambda_B$ than in 
previous work \cite{Beneke:2011nf,Wang:2016qii}.

The outline of the paper is as follows. In Sec.~\ref{sec:kinematics} 
we provide the relevant definitions, kinematics and notation. The 
subsequent two Secs.~\ref{sec:ht} and \ref{sec:soft} contain the 
results for the power-suppressed hard-collinear contributions to the 
form factor and the dispersive representation of the soft endpoint 
contributions, respectively. Sec.~\ref{sec:N} presents the numerical 
analysis of the 
form factors including the above results in several $B$-meson LCDA 
models. We summarize in Sec.~\ref{sec:summary}. Appendix~\ref{App:DAs} 
collects formulae for and relations between the two- and three-particle 
$B$-meson LCDAs up to twist four employed in this work. 

%
\section{Kinematics and notation}
\label{sec:kinematics}
%

The radiative leptonic $B$-meson decay amplitude\footnote{In the following, 
$\ell$ may refer to the electron or muon. The muon mass is set to zero 
in the kinematic expressions below.} 
\begin{equation}
A(B^-\to \gamma \ell \bar{\nu}_\ell)=\frac{G_FV_{ub}}{\sqrt{2}}
\langle \ell\bar{\nu}_l\gamma | \bar{\ell}\,\gamma^\nu(1-\gamma_5)\nu_\ell
\bar{u}\gamma_\nu(1-\gamma_5)b |B^-\rangle\,
\label{eq:defampl}
\end{equation}
can be written in terms of two form factors, $F_V$ and $F_A$,
defined through the Lorentz decomposition of the hadronic tensor
\begin{eqnarray}
T_{\mu\nu}(p,q)&=&-i\int d^4x\,e^{ipx}\langle 0 | 
T\{ j^{em}_\mu(x)\,\bar{u}(0)\gamma_\nu(1-\gamma_5)b(0)\}| B^- (p+q)\rangle
\nonumber\\&=&
\epsilon_{\mu\nu\tau \rho }p^\tau v^\rho  F_V 
+i\big[-g_{\mu\nu}(pv)+v_\mu p_\nu \big ]F_A 
-i\frac{v_\mu v_\nu}{(pv)} f_B m_B + \mbox{$p_\mu$-terms}\,.
\label{eq:Tmunu}
\end{eqnarray}
Here $p$ and $q$  are the photon and lepton-pair momenta, respectively,
so that $p+q= m_B v$ is the $B$-meson momentum in terms of its four-velocity.
In the above  $j_{\rm em}^\mu=\sum_q e_q\bar{q}\gamma_\mu q$ is the 
electromagnetic current.
The $v_\mu v_\nu$ term is fixed by the Ward 
identity \cite{Beneke:2011nf,Khodjamirian:2001ga}
\begin{align}
   p^\mu T_{\mu\nu} = -i f_B m_B v_\nu
\end{align}
and the terms proportional to $p_\mu$ contract to zero with the photon 
polarization vector, see~\cite{Beneke:2011nf} for more details. 

The form factors can be written as functions of the lepton-pair invariant 
mass squared $q^2$, or, equivalently,
of the photon energy $E_\gamma = vp $ in the $B$-meson rest frame:
\begin{align}
  q^2 = (m_Bv-p)^2 = m_B^2 + p^2 -2m_B E_\gamma\,.
\end{align}
For a real photon, $p^2=0$ and 
\begin{align}
E_\gamma = \frac{m_B^2-q^2}{2m_B}\,, 
&& 0 \leq E_\gamma \leq \frac{m_B}{2}\,, 
&& 0 \leq q^2 \leq m_B^2\,.
\end{align}
The differential decay width is given by
\begin{equation}
 \frac{d\Gamma}{dE_\gamma} =
\frac{\alpha_{\rm em}G_F^2|V_{ub}|^2}{6\pi^2}m_B E_\gamma^3
\left(1-\frac{2E_\gamma}{m_B}\right)
\left(\,\Big|F_V\Big|^2+\Big|F_A + \frac{e_\ell f_B}{E_\gamma}\Big|^2\,
\right)\,,
\label{eq:width}
\end{equation}
where, following~\cite{Beneke:2011nf},\footnote{Note the change of 
notation: $F_A$ is denoted by $\hat{F}_A$ in~\cite{Beneke:2011nf}.} 
the contact term in (\ref{eq:Tmunu}) is included in the axial form factor.

\begin{figure}[t]
\centerline{\includegraphics[width=0.3\linewidth, clip = true]{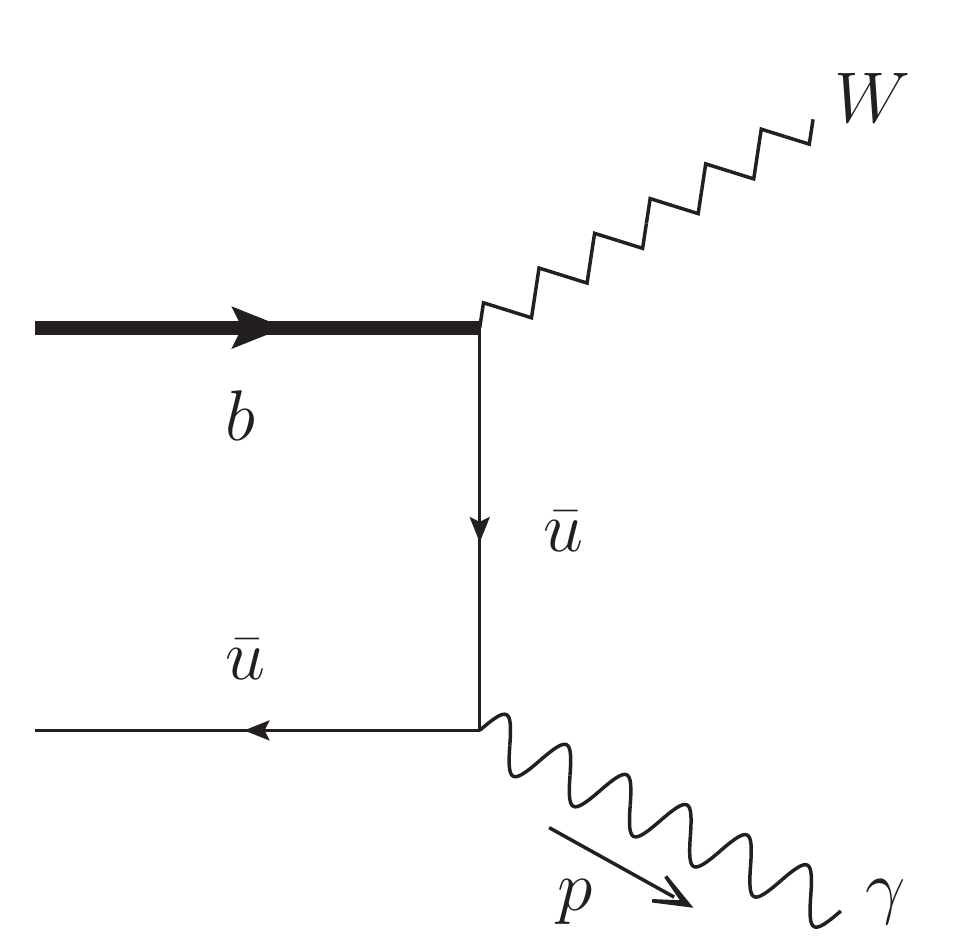}}
\caption{Leading contribution to $B\to \gamma\ell\nu_\ell$.}
\label{fig:leading}
\end{figure}

For large photon energies the form factors can be written
 as~\cite{Beneke:2011nf}
\begin{eqnarray}
F_V(E_\gamma) &=& 
\frac{e_u f_B m_B}{2 E_\gamma \lambda_B(\mu)} R(E_\gamma,\mu) 
+  \xi(E_\gamma)  +  \Delta\xi(E_\gamma)\,,
\nonumber\\
F_A(E_\gamma) &=& 
\frac{e_u f_B m_B}{2 E_\gamma \lambda_B(\mu)} R(E_\gamma,\mu) 
+  \xi(E_\gamma) - \Delta\xi(E_\gamma)\,.
\label{eq:FFs}
\end{eqnarray}
The first term is equal in both expressions and represents the leading-power  
contribution in the heavy-quark expansion (HQE). It originates only from 
photon emission from the light spectator quark in $B$ meson 
(Fig.~\ref{fig:leading}). In the above, $f_B$ is the decay constant of 
$B$ meson, and
the quantity $\lambda_B$  is the first inverse moment of the $B$-meson LCDA,
\begin{equation}
\frac{1}{\lambda_B(\mu)} = \int_0^\infty\frac{d\omega}{\omega}
\,\phi_+(\omega,\mu)\,.
\label{eq:lambdaB}
\end{equation}
The factor $R(E_\gamma,\mu)$  in (\ref{eq:FFs}) takes into account 
radiative corrections (see \cite{Beneke:2011nf} for details) and equals 
one at the tree level.

The remaining terms in~(\ref{eq:FFs}) are the power-suppressed, $1/m_b$ 
and $1/(2E_\gamma)$, corrections. They are written as a sum of the 
``symmetry-preserving'' part, i.e. the same for the both form factors 
$F_V$ and $F_A$, and the ``symmetry-breaking'' part which has
opposite sign. The leading contributions to the symmetry-breaking part  
are~\cite{Beneke:2011nf}
\begin{align}
\Delta\xi(E_\gamma) &= \frac{e_b f_B m_B}{2 E_\gamma m_b}
+\frac{e_u f_B m_B}{(2 E_\gamma)^2}\,.
\label{eq:symbreleading}
\end{align} 

The equality of the two form factors at leading power in the heavy-quark 
and large photon energy ($E_\gamma\sim m_b$) expansion is a consequence 
of the left-handedness of the weak interaction current and 
helicity-conservation of the quark-gluon interaction in the high-energy 
limit. In terms of the helicity form factors $F_\mp \equiv 
(F_V\pm F_A)/2$, the above implies that $F_+ = \Delta\xi$ vanishes 
at leading power, while $\xi$ represents the power correction to the 
non-vanishing helicity form factor $F_-$. 
Our aim is to provide improved estimates of $\xi(E_\gamma)$ and 
$\Delta\xi(E_\gamma)$, for which currently factorization formulae 
are not available. We split the calculation into ``higher-twist 
corrections'' of order $\Lambda/E_\gamma$ and $\Lambda/m_b$ from the 
region where the currents in~\eqref{eq:Tmunu} are separated by 
a small light-cone distance $x^2 \sim 1/(m_b\Lambda)$, 
and the soft or endpoint corrections. In case of the former, 
the virtuality of the quark propagator that joins the 
weak and electromagnetic vertex (see Figure~\ref{fig:leading}) is 
hard-collinear, that is of order $m_b \Lambda$. 
The latter arise when the light-cone projection $\omega$ of the 
spectator-quark momentum become anomalously small $\omega\ll \Lambda$, 
such that the quark propagator virtuality $E_\gamma\omega$ enters 
the soft region, $\Lambda^2$. 

%
\section{Higher-twist corrections}
\label{sec:ht}
%

The higher-twist corrections can be accessed via the light-cone expansion 
of the weak--electromagnetic current product at $x^2\to 0$.

At tree level one can replace the $b$-quark by the effective heavy quark field
\begin{align}
\bar q \gamma_\mu b &= \bar q \gamma_\mu h_v 
+ \frac{1}{2m_b}\bar q \gamma_\mu i \slashed{\vec{D}} h_v +\ldots
\label{eq:HQEcurrent}
\end{align}
where $\slashed{\vec{D}} = \slashed{D} - (v\cdot D) \slashed{v}$.
Then for the $u$-quark contribution (Figure~\ref{fig:leading}) we get
\begin{eqnarray}
T_{\mu\nu}^{(u)}(p,q) &=&
-ie_u\sqrt{m_B} \int d^4x\,e^{ipx}\langle 0 | T\{ \bar u(x) \gamma_\mu u(x)\,
\bar{u}(0)\gamma_\nu(1-\gamma_5)h_v(0)\}| B(v)\rangle
\nonumber\\ 
&& \hspace*{-1cm}
-\,\frac{ie_u\sqrt{m_B}}{2m_b} \int d^4x\,e^{ipx}\langle 0 | T\{ \bar u(x) 
\gamma_\mu u(x)\,\bar{u}(0)\gamma_\nu(1-\gamma_5)
i \slashed{\vec{D}} h_v(0)\}| B(v)\rangle + \ldots \qquad
\label{eq:HQE}
\end{eqnarray}
up to terms $\mathcal{O}(1/m_b^2)$ which will be neglected. 

We should note that (\ref{eq:HQEcurrent}) 
does not represent the correct heavy-quark/large-energy expansion of 
the weak current when the covariant derivative contains a collinear 
gluon field. In this case the tree-level expansion of the current 
in soft-collinear effective theory must be used (see \cite{Beneke:2002ni}). 
However, we shall make use of the above expression only to compute the 
light-cone expansion of the $u$-quark propagator 
in a soft-gluon background. Whenever a diagram 
involves a hard-collinear gluon propagator (as is the case for 
radiative corrections and the soft factorizable four-particle contribution 
discussed in the following section), we compute the corresponding 
contribution with the QCD current and quark-gluon vertex rather than 
the effective ones. 

For the contribution in the first line of (\ref{eq:HQE}), 
using the light-cone expansion of the quark propagator~\cite{Balitsky:1987bk}
\begin{eqnarray}
\wick{1}{<1 u(x) >1 {\overline{u}}(0)} &=&  
\frac{i}{2\pi^2} \frac{\slashed{x}}{x^4}
-\frac{1}{8\pi^2x^2}
\int_0^1du \bigg\{ix^\rho g\widetilde G_{\rho\sigma}(ux)\gamma^\sigma\gamma_5 
+ (2u-1) x^\rho gG_{\rho\sigma}(ux)\gamma^\sigma\biggr\}  +\ldots\,,
\nonumber\\[-0.1cm]
\\[-0.8cm]
\nonumber
\end{eqnarray}
and the definitions of $B$-meson LCDAs collected in Appendix~\ref{App:DAs} 
we obtain
\begin{eqnarray}
T_{\mu\nu}^{(u)}&=&
\frac{ie_u f_B m_B}{2\pi^2} \int\! d^4x\,\frac{e^{ipx}}{x^4}\,
\big[(vx) g_{\mu\nu} + i \epsilon_{\mu\nu\rho\sigma}x^\rho v^\sigma\big]
\nonumber\\ 
&&\times\, \biggl\{\Phi_+(vx) + x^2 G_+(vx) - \frac{x^2}{4} \int_0^1du\, 
\Big[(2u-1) \Psi_4 - \widetilde{\Psi}_4\Big](vx,uvx)
\biggr\}+\ldots
\nonumber\\
&=& 
\frac{ie_u f_B m_B}{2\pi^2} \int\! d^4x\,\frac{e^{ipx}}{x^4}\,
\big[(vx) g_{\mu\nu} + i \epsilon_{\mu\nu\rho\sigma}x^\rho v^\sigma \big]
\nonumber \\
&&\times\,
\biggl\{\Phi_+(vx) + x^2 G^{\rm\scriptscriptstyle WW}_+(vx) 
- \frac{x^2}{2(vx)^2}\Phi_-^{\rm t3}(vx) 
- \frac{x^2}{4} \int_0^1du\, \Big[\Psi_4 
- \widetilde{\Psi}_4\Big]^{\rm t4}(vx,uvx)
\biggr\}\nonumber \\[0.2cm]
&&+\,\ldots
\label{eq:Tmn}
\end{eqnarray}
where in the second representation we combined the ``genuine'' higher-twist 
contributions to the two-particle DA $G_+$ with the contributions of 
three-particle LCDAs. The ellipses stand for the contributions 
proportional to 
$p_\mu (p_\nu)$, $v_\mu (v_\nu)$ and terms $\mathcal{O}(1/E_\gamma^3)$. 
Note that the first term $\Phi_+(vx)$ in the curly bracket produces
not only the known leading-order ($R(E_\gamma,\mu) \to 1$ in (\ref{eq:FFs})),  
leading-power expression that we drop here,
but also a power correction that has to be retained.

Going over to the momentum space and identifying the two relevant Lorentz 
structures, we find from this expression
\begin{align}
\xi^{\rm ht}_{\scriptscriptstyle 1/E_\gamma} 
&=  
\frac{e_uf_B m_B}{4 E_\gamma^2}\biggl\{ 1 - 2 \frac{\bar\Lambda}{\lambda_B} 
+ 2 \int_0^\infty\!\!\!{d\omega}\, \ln \omega\, \phi_-^{\rm t3}(\omega)
\notag\\[0.1cm]
&\hspace*{2.25cm} + \int_0^\infty\!\frac{d\omega_1}{\omega_1 }\! 
\int_0^\infty\!\!\frac{d\omega_2}{\omega_1\!+\!\omega_2}\,
\big[\psi_4\! -\! \widetilde{\psi}_4\big]^{\rm t4}(\omega_1,\omega_2) \biggr\}
\notag\\[0.1cm]
&=  
\frac{e_uf_B m_B}{4 E_\gamma^2}\biggl\{ 
- 1+  2 \int_0^\infty\!\!\!{d\omega}\, \ln \omega\, \phi_-^{\rm t3}(\omega)
- \int_0^\infty\!\frac{d\omega_1}{\omega_1 }\! \int_0^\infty\!
\frac{d\omega_2}{\omega_2}\,
\big[\psi_4\! +\! \widetilde{\psi}_4\big](\omega_1,\omega_2)
\biggr\}
\notag\\[0.1cm]
&=  
\frac{e_uf_B m_B}{4 E_\gamma^2}\biggl\{ 
- 1+  2 \int_0^\infty\!\!{d\omega}\, \ln \omega\, \phi_-^{\rm t3}(\omega)
- 2 \int_0^\infty\!\frac{d\omega_2}{\omega_2}\,\phi_4(0,\omega_2) \biggr\},
\label{eq:ht1xi}
\\[0.2cm]
\Delta \xi^{\rm ht}_{\scriptscriptstyle 1/E_\gamma}&=  
\frac{e_uf_B m_B}{4 E_\gamma^2}\,.
\label{eq:ht1dxi}
\end{align}
Interestingly, $\xi^{\rm ht}_{\scriptscriptstyle 1/E_\gamma}$ only involves 
the three-particle LCDAs on the line $x=0$ in the 
$(s,x)$-representation~\cite{Braun:2017liq}, see~\eqref{sx-rep},
that are directly related by the equation of motion (EOM) 
relations~\eqref{EOM}, \eqref{EOM2}.  This property allows one to rewrite 
the answer in several equivalent forms, as shown above. 
The $\Delta \xi^{\rm ht}_{\scriptscriptstyle 1/E_\gamma}$-term arises 
from the  first term $\Phi_+(vx)$ in the curly bracket and agrees 
with the corresponding contribution in (\ref{eq:symbreleading})  
obtained in~\cite{Beneke:2011nf} 
by a different method.

From the analysis of the renormalization group behaviour~\cite{Braun:2017liq} 
one expects 
\begin{eqnarray}
&& \phi_-^{\rm t3}(\omega)\sim \omega^0, \qquad
\phi_4(0,\omega_2)\sim \omega_2^2,\qquad
 g^{\rm\scriptscriptstyle WW}_+(\omega) \sim \omega^2\,,
\nonumber\\[0.1cm] 
&& \phi_3(\omega_1, \omega_2)\sim \omega_1\omega_2^2, \qquad
\psi_4(\omega_1,\omega_2) \sim \widetilde{\psi}_4(\omega_1,\omega_2) 
\sim \omega_1 \omega_2\,,  
\end{eqnarray} 
so that all integrals are endpoint-finite for small $\omega_i$. Hence, 
to the twist-four accuracy there is no overlap with the soft region. 
Contributions of higher twist-five, six, etc., are suppressed  
by extra powers of the photon energy $E_\gamma$ in the hard-collinear 
region. These contributions can, however, have power-like endpoint 
divergences that spoil the power counting. Hence, the soft contributions 
from higher-twist terms are not necessarily suppressed by powers 
of $1/E_\gamma$ relative to the twist-four terms. 
We will discuss this mechanism in more detail in the 
next section.

The contribution from the second line in~\eqref{eq:HQE} can be calculated 
using the operator identity
\begin{align}
\bar q(x) \Gamma \derright_\xi h_v(0) &=
 {\partial_\xi} \bar q(x) \Gamma h_v(0)
+ i\int_0^1\! du\,\bar u\, \bar q(x) x^\rho gG_{\rho\xi}(ux) \Gamma h_v(0)
-   \frac{\partial}{\partial x^\xi} \bar q(x) \Gamma h_v(0)\,.
\end{align}
Since this contribution is suppressed by $1/m_b$, for this case we only 
need the leading term in the  $1/E_\gamma$, expansion. We obtain
\begin{align}
\xi^{\rm ht}_{\scriptscriptstyle 1/m_b} &=
\frac{e_u f_B m_B}{4 m_bE_\gamma}\biggl\{
\frac{\bar\Lambda }{\lambda_B} -2  
+ \int_0^\infty \! d\omega\, \ln \omega\, \phi_-^{\rm t3}(w)
\notag\\[0.1cm]&~
+ 2  \int_0^\infty \!\frac{d\omega_1}{\omega_1}  \int_0^\infty 
\!\frac{d\omega_2}{\omega_2}\,
{\phi}_3(\omega_1,\omega_2) \,\Big\{1  - \frac{\omega_1}{\omega_2} 
\ln \frac{\omega_1+\omega_2}{\omega_1}\Big\}
      \biggr\},
\notag\\[0.1cm]&= 
  \frac{e_u f_B m_B}{4 m_bE_\gamma}\biggl\{
\frac{\bar\Lambda }{\lambda_B} -2  
+ 2  \int_0^\infty \!\frac{d\omega_1}{\omega_1}  \int_0^\infty 
\!\frac{d\omega_2}{\omega_1+\omega_2}\,
{\phi}_3(\omega_1,\omega_2)
\biggr\},
\label{eq:htmb2}
\\[0.2cm]
  \Delta  \xi^{\rm ht}_{\scriptscriptstyle 1/m_b} &=0\,, 
\label{eq:htmbd2}
\end{align}
where $\phi_-^{\rm t3}(\omega)$ is the ``genuine'' twist-three contribution 
to the LCDA $\phi_-(\omega)$, cf.~\eqref{Phi+Phi-}. 

The complete higher-twist corrections $1/E_\gamma, 1/m_b$ are given by the 
sum of the above two contributions
\begin{align}
\xi^{\rm ht} &=   \xi^{\rm ht}_{\scriptscriptstyle 1/E_\gamma}
+  \xi^{\rm ht}_{\scriptscriptstyle 1/m_b},
\notag\\[0.1cm]
    \Delta\xi^{\rm ht}  &= \frac{e_b f_B m_B}{2 E_\gamma m_b}
+\frac{e_u f_B m_B}{(2 E_\gamma)^2}\,,
    \label{eq:HTcorrection}
\end{align}
where for $\Delta\xi^{\rm ht}$ we have added the contribution of the photon 
emission from the $b$-quark~\cite{Beneke:2011nf}. The second equation 
of~(\ref{eq:HTcorrection}) 
agrees with the previous result (\ref{eq:symbreleading}). 
The absence of an endpoint divergence in the higher-twist 
correction arises as a consequence of non-trivial relations between 
the various terms in (\ref{eq:Tmn}) and it would be interesting 
to understand this in the context of a factorization theorem for 
the $1/E_\gamma$ power corrections, which is, however, beyond the scope 
of the present paper. A previous attempt~\cite{Wang:2016qii} to compute 
$\xi^{\rm ht}_{\scriptscriptstyle 1/E_\gamma}$ did not include the 
$G_+$ term and used an incorrect parametrization of the three-particle 
matrix element (\ref{def:three}), resulting in a qualitatively 
different, endpoint-divergent higher-twist correction.

%
\section{Soft corrections}
\label{sec:soft}
%

In addition to higher-twist corrections, the power-suppressed contributions 
to the form factors can originate from large distances between the currents 
in~\eqref{eq:Tmunu}, $x^2 \sim 1/\Lambda^2$, which cannot be 
accessed in the light-cone expansion. Such contributions may or may not be 
``visible'' through the infrared (endpoint) divergences of the 
hard-collinear higher-twist contributions, and cannot be factorized in terms 
of the LCDAs without additional assumptions. We will use the approach 
suggested in \cite{Braun:2012kp} that is based on using dispersion relations 
and quark-hadron duality. This technique has originally been proposed for 
the study of the $\gamma^*\gamma\to \pi$ transition form 
factor~\cite{Khodjamirian:1997tk} and has become the method of choice for 
this reaction, see e.g.~\cite{Agaev:2010aq,Mikhailov:2016klg} for 
recent refinements. Our aim is to put the calculation of the 
radiative leptonic decay form factors on the same level as the  
$\gamma^*\gamma\to \pi$ transition form factor.

The starting point is the more general process $B\to \gamma^* \ell \nu_\ell$ 
with a transversely polarized, virtual photon with $p^2<0$. If 
$-p^2 \sim m_B \Lambda$, the correlation function 
in~(\ref{eq:Tmunu}) does not receive any soft contribution
and can be calculated (in principle) in terms of the $B$-meson LCDAs of 
increasing twist to arbitrary power in the $1/E_\gamma, 1/m_b, 
1/p^2$ expansion. The idea is to access the real photon limit $p^2=0$ 
starting from this expansion by using the dispersion relation.  In this way, 
the explicit evaluation of soft contributions is effectively replaced by a 
certain ansatz (assumption) for the hadronic spectral density in the 
$p^2$-channel. The procedure can be understood as the matching of two 
different representations for the correlation function~(\ref{eq:Tmunu}) --- 
the QCD calculation in terms of quarks and gluons vs. physical hadrons in the 
intermediate state --- and is usually referred to
as light-cone sum rules (LCSR)~\cite{Balitsky:1989ry}. 

On the one hand, one can argue on general grounds that the generalized form 
factors $F_{B\to \gamma^*}(E_\gamma,p^2)$  ($F_{B\to \gamma^*}$ refers to 
both,  vector and axial, $F_V$ and $ F_A$) satisfy an unsubtracted dispersion 
relation in the variable $p^2$ at fixed  
$2 m_B E_\gamma \equiv {2 m_B}\, vp = m_B^2 + p^2 -q^2$. 
Separating the contribution of the lowest-lying vector mesons 
$\rho,\omega$, we write 
\begin{equation}
F_{B\to \gamma^*}(E_\gamma,p^2)= 
\frac{f_\rho F_{B\to \rho}(q^2)}{m^2_\rho-p^2} + 
\frac{1}{\pi}\int_{s_0}^\infty ds\,
\frac{\mathrm{Im} F_{B\to \gamma^*}(E_\gamma,s)}{s-p^2}\,,
\label{eq:DR}
\end{equation}
where $s_0$ defines an effective 
continuum threshold. For simplicity we combined here the $\rho$ and $\omega$ 
contributions in one resonance term assuming $m_\rho\simeq m_\omega$ and the 
zero-width approximation. In this expression, $f_\rho$  is the usual decay 
constant of the vector meson and $F_{B\to \rho}(q^2)$ is a generic 
$B\to \rho(\omega)$ transition form factor, whose explicit definition 
will not be needed. Since there are no massless states, the real photon 
limit is recovered by setting $p^2\to 0$ in (\ref{eq:DR}).  

On the other hand, the same form factors can be calculated for sufficiently 
large $-p^2$ using QCD factorization. The result, 
$F^{\rm QCDF}_{B\to \gamma^* }(E_\gamma,p^2)$, satisfies a similar dispersion 
relation
\begin{equation}
F^{\rm QCDF}_{B\to \gamma^* }(E_\gamma,p^2) = 
\frac{1}{\pi}\int_{0}^\infty ds\,
\frac{\mathrm{Im} \,F^{\rm QCDF}_{B\to \gamma^*}(E_\gamma,s)}{s-p^2}\,,
\label{eq:DRQCD}
\end{equation}
where the limit $p^2\to 0$ cannot be taken directly. Singular terms in 
$1/p^2$ appear (cf.~\cite{Agaev:2010aq} for the case of the 
$\gamma\gamma^*\to\pi^0$ form factor), signalling that QCD factorization 
cannot be applied directly to the real photon case $p^2=0$ beyond the leading 
power in $1/m_b$ and $1/E_\gamma$.
 
The main assumption of the method (quark-hadron duality) is that the physical 
spectral density above the threshold $s_0$ coincides with the calculated in 
QCD spectral density upon averaging with a smooth weight function over a 
sufficiently broad interval of the 
energy $s$:
\begin{equation}
\mathrm{Im}\,F_{B\to \gamma^*}(E_\gamma,s) \simeq 
\mathrm{Im}\,F^{\rm QCDF}_{B\to \gamma^* }(E_\gamma,s)
\qquad \mbox{for}~~s>s_0\,.
\label{eq:duality}
\end{equation}
For the simplest sum rule, one uses that the QCD factorization calculation 
must reproduce the ``true'' form factors $F_{B\to \gamma^*}(E_\gamma,p^2)$  
for asymptotically large values of $-p^2$. Equating the two representations 
(\ref{eq:DR}) and (\ref{eq:DRQCD}) at $p^2\to-\infty$ 
and subtracting the contributions of $s>s_0$ from the both sides one obtains
\begin{eqnarray}
f_\rho F_{B\to \rho }(q^2) = \frac{1}{\pi}\int_{0}^{s_0}ds\, 
\mathrm{Im} \,F^{\rm QCDF}_{B\to \gamma^* }(E_\gamma,s)\,.
\label{eq:beforeBorel}
\end{eqnarray}
In practical applications of this method one uses an additional 
trick~\cite{Shifman:1978bx} which allows one to reduce the sensitivity 
to the duality assumption in~(\ref{eq:duality}) and simultaneously suppress 
the contributions of higher twists in the light-cone expansion.  This is done 
by passing to the Borel representation of the dispersion 
relation, which effectively substitutes $1/(s-p^2)\to \exp(-s/M^2)$. 
The net effect on (\ref{eq:beforeBorel}) is the appearance of an additional 
weight factor under the integral:     
\begin{eqnarray}
f_\rho F_{B\to \rho }(q^2) = 
\frac{1}{\pi}\int_{0}^{s_0}ds\, e^{-(s-m^2_\rho)/M^2}\,
\mathrm{Im} \,F^{\rm QCDF}_{B\to \gamma^* }(E_\gamma,s)\,.
\label{eq:Frhogamma}
\end{eqnarray}
The value of the Borel parameter $M^2$ corresponds, roughly speaking, to the inverse (Euclidean) distance at which the matching is done 
between the quark and hadron representations. In ideal case there 
should be no $M^2$-dependence so that varying  $M^2$ within a certain window, usually $M^2=1-2$~GeV$^2$, one obtains an 
indication of the accuracy of the calculation.   

With this refinement, substituting (\ref{eq:Frhogamma}) into (\ref{eq:DR}) 
and using (\ref{eq:duality}), one obtains for $p^2\to 0$~\cite{Braun:2012kp} 
\begin{align}
F_{B\to \gamma}(E_\gamma) &= 
\frac{1}{\pi}\int_{0}^{s_0} \frac{ds}{m_\rho^2} \,
\mathrm{Im} \,F^{\rm QCDF}_{B\to \gamma^* }(E_\gamma,s)\,e^{-(s-m^2_\rho)/M^2} 
+ \frac{1}{\pi}\int_{s_0}^\infty \frac{ds}{s} \,
\mathrm{Im} \,F^{\rm QCDF}_{B\to \gamma^* }(E_\gamma,s)
\notag\\[0.2cm]
&=
F^{\rm QCDF}_{B\to \gamma }(E_\gamma)  
+  \xi^{\rm soft}_{B\to \gamma}(E_\gamma)\,.
\label{eq:LCSR}
\end{align}
In passing to the second line we extend the lower limit of the 
second integral to 0 and subtract the added contribution from 
the first.  In this way the second integral equals 
$F^{\rm QCDF}_{B\to \gamma }(E_\gamma) = 
F^{\rm QCDF}_{B\to \gamma^* }(E_\gamma,p^2=0)$ calculated using QCD 
factorization and 
\begin{align}
 \xi^{\rm soft}_{B\to \gamma}(E_\gamma) = 
\frac{1}{\pi}\int_{0}^{s_0}\!\frac{ds}{s}\,
\Big[ \frac{s}{m_\rho^2}e^{-(s-m^2_\rho)/M^2}  - 1 \Big]\, 
\mathrm{Im} \,F^{\rm QCD}_{B\to \gamma^* }(E_\gamma,s)
\label{eq:xi-soft}
\end{align}
is the soft correction that originates from the nonperturbative modification 
of the spectral density. Conceptually, the effect of this modification 
is to create a mass gap in the vector-meson mass spectrum. Separating in 
\eqref{eq:xi-soft} the contributions that are the same for the form factors 
$F_V$ and $F_A$, and those of opposite sign, we can decompose the soft 
correction in the ``symmetry-preserving'' part $\xi^{\rm soft}(E_\gamma)$, and the ``symmetry-breaking'' part $\Delta\xi^{\rm soft}(E_\gamma)$
in the notation of~\eqref{eq:FFs}.
 
Note that both terms in the first line of~\eqref{eq:LCSR} and hence the full 
result are finite, whereas the decomposition as the sum of the ``pure'' QCD 
factorization expression and the soft correction in the second line can 
in principle (but not in the above) produce logarithmic and/or power 
divergences from the $s\to 0$ region. In such cases (see example below), 
for bookkeeping purposes we will attribute the whole contribution to the 
soft correction. 

In the following we apply~\eqref{eq:LCSR} to the leading-power and 
higher-twist hard-collinear contributions calculated in~\cite{Beneke:2011nf} 
and in the previous section. That is, for each hard-collinear contribution 
to $F^{\rm QCDF}_{B\to \gamma }(E_\gamma)$ we obtain the corresponding   
$\xi^{\rm soft}_{B\to \gamma}(E_\gamma)$ due to the nonperturbative 
modification of the spectral function in the soft region.

The soft correction to the leading-order, leading-twist hard-collinear 
contribution given by the first term in the two equations (\ref{eq:FFs}) 
with $R(E_\gamma,\mu)$ set to 1 was considered in~\cite{Braun:2012kp}.
For the form factors at non-vanishing $p^2$, we obtain, 
\begin{equation}
F^{(\LO)}_V(E_\gamma,p^2) =  F^{(\LO)}_A(E_\gamma,p^2) = e_u f_B m_B 
\,U_{\rm LL}
\int_0^\infty{d\omega}\,\frac{\phi_+(\omega,\mu)}{2 E_\gamma \omega - p^2}\,.
\label{eq:Fzero} 
\end{equation} 
Here $U_{\rm LL}$ is the renormalization-group factor 
$U(E_\gamma,\mu_{h1},\mu_{h2},\mu)$~\cite{Beneke:2011nf} truncated to the 
leading-logarithmic approximation,\footnote{See Appendix~A 
of~\cite{Beneke:2011nf}. In the 
leading-logarithmic approximation, the $\alpha_s(\mu_h)$ terms in (A.3) 
are neglected. We follow the terminology of~\cite{Beneke:2011nf}, 
which implies that LL includes the two-loop cusp and one-loop non-cusp 
anomalous dimension in the renormalization group equation, NLL three-loop 
cusp and two-loop non-cusp, and so on.} which sums large logarithms 
from the hard scales $\mu_{h1},\mu_{h2} \sim m_b, E_\gamma$ to a 
hard-collinear scale of order $-p^2$.
The integral in (\ref{eq:Fzero}) can easily be converted to the form of a 
dispersion relation by the change of variables $s = 2 E_\gamma \omega$. 
{}Following the procedure described above and changing the integration 
variable back to $\omega = s/(2E_\gamma)$, we obtain 
\begin{align}
\xi^{\rm soft}_{(\LO)}(E_\gamma) &= 
\frac{e_u f_B m_B}{2E_\gamma}\,U_{\rm LL}
\int_0^{\frac{s_0}{2 E_\gamma}}\!d\omega \biggl[\frac{2E_\gamma}{m^2_\rho}
e^{-(2E_\gamma\omega- m_\rho^2)/M^2}-\frac{1}{\omega}\biggr]
\phi_+(\omega,\mu)\,,
\notag\\[0.2cm]
\Delta\xi^{\rm soft}_{(\LO)}(E_\gamma) &= 0\,. 
\label{eq:xi-soft-LO}
\end{align}
The soft correction defined by~\eqref{eq:xi-soft-LO} comes from 
the region $\omega < s_0/(2E_\gamma)\sim s_0/m_b$. With $\sqrt{s_0}$ a few 
times $\Lambda$, this is indeed an endpoint spectator-quark contribution 
corresponding to contribution from a soft distance $1/\Lambda$ between 
the weak and electromagnetic current in (\ref{eq:Tmunu}).  
Since for large scales $\mu\sim m_b$ the LCDA $\phi_+(\omega,\mu) \sim 
\omega $ for $\omega\to 0$ one obtains a power correction of the order 
of $s_0/(2E_\gamma{\Lambda})$ for $E_\gamma\sim m_b \to \infty$ with respect to the 
leading, hard-collinear contribution, in agreement with the usual power 
counting for the soft form factor $\xi(E_\gamma)$. Note that since 
the shape of $\phi_+(\omega,\mu)$ is governed by the QCD scale $\Lambda$, 
while $\omega$ is restricted to values smaller than 
$s_0/(2E_\gamma)\ll \Lambda$ in \eqref{eq:xi-soft-LO}, one might be 
tempted to approximate $\phi_+(\omega,\mu)$ by its asymptotic 
behaviour $\phi_+(\omega,\mu) \sim \omega$ as $\omega\to 0$. However, 
this would amount to the first term in an expansion of the integral 
in powers of $s_0/(E_\gamma \Lambda)$, which for realistic values 
of $s_0\approx 1.5~$GeV${}^2$ and 
$E_\gamma  \sim 1.5-2.5$~GeV is not a valid approximation. We therefore 
always keep the full functional form of the LCDA in the integrals 
for the soft contributions.

Applying the same method to the next-to-leading order 
${\cal O}(\alpha_s)$ correction 
to the leading-twist contribution requires factorizing the hadronic 
tensor into a hard matching coefficient $C(E_\gamma,\mu_{h1})$ 
and a hard-collinear function, 
which is convoluted with $\phi_+(\omega)$. The hard function is independent 
of the hard-collinear variable $-p^2$ and is given in~\cite{Beneke:2011nf}. 
The hard-collinear function calculated for $p^2=0$ in   
\cite{Lunghi:2002ju,Bosch:2003fc} must be generalized to $-p^2\not=0$. 
The  result can be brought into the form~\cite{Wang:2016qii} 
\begin{eqnarray}
\xi^{\rm soft}_{(\NLO)}(E_\gamma)&=&
\frac{e_uf_Bm_B}{2E_\gamma}\,C(E_\gamma,\mu_{h1})K^{-1}(\mu_{h2}) 
U(E_\gamma,\mu_{h1},\mu_{h2},\mu)
\nonumber\\
&&\times\,
\int^{\frac{s_0}{2 E_\gamma}}_0\!\! d\omega'
\left[\frac{2E_\gamma}{m_\rho^2}\,\e^{-(2E_\gamma\omega'-m_\rho^2)/M^2}
-\frac1{\omega'}\right]\,\phi_+^{\rm eff}(\omega',\mu)\,,
\notag\\
\Delta\xi^{\rm soft}_{(\NLO)}(E_\gamma)&=&0\,,
\label{eq:xi-soft-NLO}
\end{eqnarray}
where ``NLO'' is meant to include the LO contribution and the 
prefactor includes the hard NLO matching correction and 
next-to-leading-logarithmic resummation as given in~\cite{Beneke:2011nf}. 
The convolution of the generalized hard-collinear function with 
$\phi_+(\omega,\mu)$ after applying the dispersive treatment and 
letting $p^2\to 0$ at the end, is summarized in 
\begin{eqnarray}
\phi_{+}^{\rm eff} (\omega^{\prime},\mu) 
&=& \phi_{+}(\omega^{\prime},\mu) 
+ \frac{\alpha_{s}(\mu)C_{F}} {4\pi}
\bigg\{
\Big( 
\ln^{2}\frac{\mu^{2}}{2 E_\gamma\omega^{\prime}}
+ \frac{\pi^{2}}{6} - 1 
\Big)\,
\phi_{+}(\omega^{\prime},\mu)
\nonumber \\
&& + \,
\Big(2\ln\frac{\mu^{2}}{2 E_\gamma\omega^{\prime}} + 3 \Big) 
\,\omega^{\prime}
\int^{\infty}_{\omega^{\prime}} d\omega 
\ln\frac{\omega-\omega^{\prime}}{\omega^{\prime}}
\frac{d}{d\omega} \frac{\phi_{+}(\omega,\mu)}{\omega}
\nonumber \\
&& - \,2\ln\frac{\mu^{2}}{2 E_\gamma\omega^{\prime}}
\int^{\omega^{\prime}}_{0} d\omega 
\ln\frac{\omega^{\prime}-\omega}{\omega^{\prime}} 
\frac{d}{d\omega} \phi_{+}(\omega,\mu)
\nonumber\\
&& +\,\int^{\omega^{\prime}}_{0}d\omega
\ln^{2}\frac{\omega^{\prime}-\omega}{\omega^{\prime}}
\frac{d}{d\omega}
\Big[
\frac{\omega^{\prime}}{\omega} \phi_{+}(\omega,\mu) 
+ \phi_{+}(\omega,\mu)
\Big]
\bigg\}.
\label{eq:phieff}
\end{eqnarray}
The hard-collinear NLO contribution~\cite{Beneke:2011nf} can 
be written in this notation as
\begin{align}
\frac{J(E_\gamma,\mu)}{\lambda_B(\mu)}
= \int^\infty_0\frac{d\omega'}{\omega'}\,\phi_+^{\text{eff}}(\omega',\mu)\,.
\end{align}
The soft correction for the ${\cal O}(\alpha_s)$ leading-twist contribution 
was previously calculated in~\cite{Wang:2016qii}. We find that the above 
expression (\ref{eq:phieff}) can be rewritten into 
the one given in~\cite{Wang:2016qii} up to several differences that 
seem to be obvious misprints. 

For the higher-twist contributions considered in Sec.~\ref{sec:ht}, the 
dispersive treatment of the soft contribution corresponding to the 
hard-collinear terms (\ref{eq:ht1xi}), (\ref{eq:ht1dxi}), 
(\ref{eq:htmb2}) and (\ref{eq:htmbd2}) yields 
\begin{align}
\xi^{\rm soft}_{(tw-3,4)}(E_\gamma)&=
\frac{e_um_Bf_B}{4E_\gamma^2}\int^{\frac{s_0}{2E_\gamma}}_0d\omega
\left[\frac{2E_\gamma}{m_\rho^2}\,\e^{-(2E_\gamma\omega-m_\rho^2)/M^2}
-\frac1{\omega}\right]
\Xi_1(\omega)
\notag\\
&\quad +\frac{e_um_Bf_B}{4m_bE_\gamma}\int^{\frac{s_0}{2E_\gamma}}_0d\omega
\left[\frac{2E_\gamma}{m_\rho^2}\,\e^{-(2E_\gamma\omega-m_\rho^2)/M^2}
-\frac1{\omega}\right]
\Xi_2(\omega)\,,
\label{eq:xi-soft-t34pre}
\\[0.2cm]
\Delta\xi^{\rm soft}_{(tw-3,4)}(E_\gamma)&=
\frac{e_um_Bf_B}{4E_\gamma^2}\int^{\frac{s_0}{2E_\gamma}}_0d\omega
\left[\frac{2E_\gamma}{m_\rho^2}\,\e^{-(2E_\gamma\omega-m_\rho^2)/M^2}
-\frac1{\omega}\right]\omega\,\phi_+(\omega)\,.
\label{eq:xi-soft-t34bre}
\end{align} 
As was the case with \eqref{eq:ht1xi}, \eqref{eq:htmb2} the result can 
be written in several equivalent ways using the EOM, e.g.,   
\begin{eqnarray}
\Xi_1(\omega) &=&
- \int^{\omega}_0d\omega_1\int^\infty_{\omega\!-\!\omega_1}
\frac{d\omega_2}{\omega_2}
\frac \partial{\partial\omega_1}\big[\psi_4\!+\widetilde\psi_4\big]
(\omega_1,\omega_2)
\nonumber\\
&& \, - \int^{\omega}_0d\omega_2\int^\infty_{\omega\!-\!\omega_2}
\frac{d\omega_1}{\omega_1}
\frac \partial{\partial\omega_2}\big[\psi_4+\widetilde\psi_4\big] 
(\omega_1,\omega_2)
\nonumber \\
&&
+\, 2 \int_{\omega}^\infty\! d\rho\,\phi_-^{\rm t3}(\rho) - 
2 \omega  \phi_-^{\WW}(\omega) + 2 \omega \phi_+(\omega) 
+ \omega^2 \frac{d}{d\omega}\phi_+(\omega)\,,
\label{Xi1}\\[0.3cm]
\Xi_2(\omega) &=& 
2 \int^\infty_0\frac{d\omega_2}{\omega_2}\,\phi_3(\omega,\omega_2)
- 2 \int^{\omega}_0d\omega_1\int^\infty_{\omega\!-\!\omega_1}
\frac{d\omega_2}{\omega_2^2}\,\phi_3(\omega_1,\omega_2)
+\int_{\omega}^\infty\!d\rho\,\phi_-^{\rm t3}(\rho)
\notag\\[0.1cm]
&&
+ \,(\bar\Lambda-\omega)\,\phi_+(\omega)-\omega\phi_-^{\WW}(\omega)\,. 
\label{Xi2}
\end{eqnarray}
In these expressions we used 
that $\big[\psi_4\!+\!\widetilde\psi_4\big] (\omega_1=0,\omega_2) 
= \big[\psi_4+\widetilde\psi_4\big] (\omega_1,\omega_2=0) = 0$.

Since the higher-twist contributions in Sec.~\ref{sec:ht} do not suffer 
from a soft endpoint divergence for real photon emission, the modification 
of the spectral density according to~\eqref{eq:LCSR} results in a soft 
correction (\ref{eq:xi-soft-t34pre}), which is suppressed by an additional 
power of $E_\gamma$ and is therefore, strictly speaking, beyond our accuracy. 
However, the actual suppression factor relative to the leading-power 
form factor is $\{1/E_\gamma^2,1/(m_b E_\gamma)\}\times 
s_0/(E_\gamma \Lambda)$ 
and since $s_0/(E_\gamma\Lambda) \gg \Lambda/E_\gamma$ such 
corrections can be numerically significant. We recall that also for the 
leading-twist contributions \eqref{eq:xi-soft-LO} and \eqref{eq:xi-soft-NLO}, 
we keep the full expressions and do not expand the result in powers of 
$s_0/(E_\gamma \Lambda)$, $M^2/(E_\gamma\Lambda)$, since this expansion
converges very slowly for realistic energies $E_\gamma \sim 1.5-2.5$~GeV.      
Thus we take the soft corrections due to twist-three and twist-four 
contributions into account in the numerical analysis.\footnote{As noted 
above, the dispersion relation is done in $p^2$ at fixed 
$E_\gamma \equiv vp$. Using instead the ``canonical'' dispersion relation 
in $p^2$  for fixed $q^2$ would lead to the following modifications: 
First, an extra factor $(1-\omega/m_B)$ appears in the 
denominator of the $\omega$-integral in (\ref{eq:Fzero}). Second, the upper 
limit on the invariant mass is redefined from $2 E_\gamma \omega - \omega^2 < 
s_0$ to $2 E_\gamma \omega - \omega^2 < s_0 \,(1-\omega/m_B)$. In both 
versions we expand this constraint assuming $\omega \sim \Lambda_{\rm QCD} 
\ll E_\gamma, m_b$ and take into account  $\mathcal{O}(\omega/E_\gamma)$, 
$\mathcal{O}(\omega/m_B)$ terms, which is consistent with twist-four accuracy 
in the collinear expansion. The resulting difference in the soft correction, 
which is already suppressed as $1/E_\gamma$ with respect to the leading term,
is suppressed by an additional factor $1/m_B$, and should be viewed as an 
ambiguity of the method. This ambiguity is, in principle, of the same order 
as the term in $\Xi_2$ in \eqref{eq:xi-soft-t34pre}. For the other terms it 
is yet higher order, if we accept that terms $(s_0/(E_\gamma \Lambda))^k$
should be retained whenever possible, whereas higher-order  terms in 
$s_0/(m_b \Lambda)$ can be dropped. We note that different terms in $\Xi_1$ 
formally contribute at different order in the $s_0/(E_\gamma \Lambda)$ 
expansion; e.g. the  term $\omega \phi_+(\omega)$ in the expression for 
$\Xi_1$ contributes to the form factor only at order 
$1/E_\gamma^2 \,(s_0/(E_\gamma \Lambda))^2$. The symmetry-breaking soft 
contribution (\ref{eq:xi-soft-t34bre}) is entirely of this order.}

We stress again that the soft contribution cannot be obtained through the 
light-cone expansion of the current product and as a consequence the usual 
hierarchy of the contributions of different (collinear) twist breaks 
down: $B$-meson LCDAs of all twists can in principle contribute to the 
form factor to the same power $1/E_\gamma^2$ in the $1/E_\gamma$ expansion. 
The basic idea of the light-cone sum rule approach is that higher-twist LCDAs 
have higher dimension and their contribution to the form factors is, 
generically, suppressed by increasing powers of the Borel parameter or 
continuum threshold, $s_0, M^2 \gg \Lambda^2$. Thus one can hope that only 
the first few terms in this expansion are numerically important.

\begin{figure}[t]
\centerline{\includegraphics[width=0.8\linewidth]{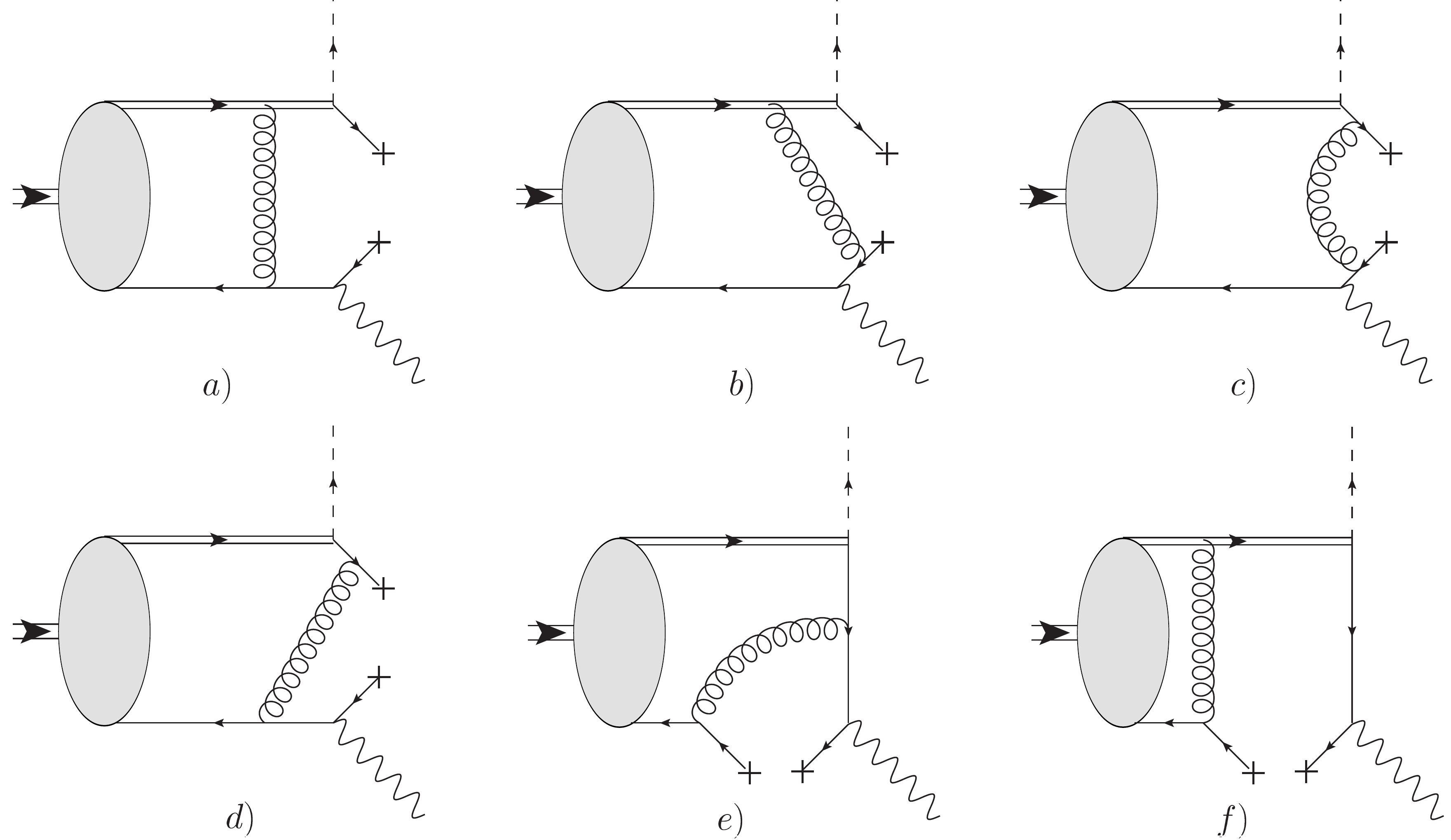}}
\caption{Factorizable higher-twist corrections to $B\to \gamma\ell\nu_\ell$}
\label{fig:factor}
\end{figure}

As noted above, this does not actually happen for the twist-3 and -4 
contributions, at least at the tree-level, due to their endpoint finiteness. 
In the following we consider the simplest higher-twist contribution of 
this unsuppressed kind, the contribution of twist-five and twist-six 
four-particle LCDAs in the factorization approximation --- as a product of 
lower-twist LCDAs and the quark condensate (cf.~\cite{Agaev:2010aq}), see 
the relevant diagrams in Fig.~\ref{fig:factor}. For the diagram 
in Fig.~\ref{fig:factor}~a we obtain after a short calculation 
\begin{eqnarray}
T_{\mu\nu}^{\Figa}(p,q) &=&
\frac{ie_u f_B m_B g_s^2 C_F \langle \bar u u\rangle}{48 p^2 E_\gamma}
 \Tr\Big[\slashed{p}\gamma_\mu\gamma_\nu (1-\gamma_5) \slashed{v} \Big] 
\,\int_0^\infty\!d\omega 
\frac{\phi_-(\omega)}{p^2-2E_\gamma \omega} 
\nonumber\\
&& +\,\mathcal{O}(1/E_\gamma^3)\,, 
\label{eq:fig1a}
\end{eqnarray}
where $\langle \bar u u\rangle \approx -(240\,\text{MeV})^3$ (at the scale 
1~GeV) is the quark condensate. If $|p^2| \sim E_\gamma \Lambda$ this 
contribution is suppressed by three powers of the hard-collinear scale in 
agreement with twist counting. However, the real photon limit $p^2\to 0$ 
cannot be taken because of the $1/p^2$ factor, and also the integral of the 
twist-three DA $\phi_-(\omega)$ becomes logarithmically divergent in this 
limit. The effect of the dispersion relation improvement  
is, for the simplest case of a pure pole in $p^2$, the 
substitution~\cite{Balitsky:1985aq}
\begin{align}
\frac{1}{-p^2} \mapsto \frac{1}{m_\rho^2-p^2} 
\stackrel{p^2\to 0}{\longrightarrow} \frac{1}{m_\rho^2}\,.
\end{align}
In this way the contribution to the form factors corresponding to 
\eqref{eq:fig1a} remains finite but the power counting changes and we obtain 
a term $\mathcal{O}(1/E^2_\gamma)$ similar to the hard-collinear contribution 
of the twist-three and twist-four LCDAs considered in Sec.~\ref{sec:ht}. 

The other diagrams in Fig.~\ref{fig:factor} can be evaluated in a similar 
manner.\footnote{The calculation of the diagrams in in 
Figs.~\ref{fig:factor}~b,~c,~d is straightforward, while 
Fig.~\ref{fig:factor}~e can most easily be obtained using the 
background-field expansion of the quark propagator~\cite{Balitsky:1987bk}.
Fig.~\ref{fig:factor}~f effectively corresponds to a contribution from  
the two-particle twist-five LCDA $g_-(\omega)$ (see App.~\ref{App:DAs}), 
which can be factorized into a product of the quark condensate and a 
lower-twist LCDA. Figs.~\ref{fig:factor}~a,~b involve a hard-collinear gluon 
propagator and therefore have to be calculated with the full QCD current 
and vertex, as mentioned before. It turns out, however, that the difference 
to using HQET rules appears only at order, $1/(m_b E_\gamma^2)$, 
beyond the accuracy of our calculation.}
We find that the contributions in Figs.~\ref{fig:factor}~a,~c,~e get promoted 
in the limit $p^2\to 0$ to a $1/E_\gamma^2$ correction, whereas the 
contributions in Figs.~\ref{fig:factor}~b,~d,~f remain of order 
$\mathcal{O}(1/E_\gamma^3)$ and can be neglected. We obtain
\begin{align}
\xi^{\rm soft}_{(tw-5,6)}(E_\gamma)&=
\frac{e_ug_s^2C_F\vev{\bar uu}f_B m_B}{48E_\gamma^2\,m_\rho^2}
\biggl\{\e^{m_\rho^2/M^2} \int^{\frac{s_0}{2 E_\gamma}}_0
\frac{d\omega}{\omega}\Big(\e^{-2E_\gamma\omega/M^2}-1\Big)\phi^{\WW}_-(\omega)
\notag\\
&\quad+\int^\infty_{\frac{s_0}{2 E_\gamma}}\frac{d\omega}{\omega}
\Big(\frac{m_\rho^2}{2E_\gamma\omega}-\e^{m_\rho^2/M^2} \Big)\phi^{\WW}_
-(\omega)-\frac5{\lambda_B}\e^{m_\rho^2/M^2}\biggr\}\, ,
\notag\\[0.2cm]
\Delta\xi^{\rm soft}_{(tw-5,6)}(E_\gamma)&=
-\frac{e_ug_s^2C_F\vev{\bar uu}f_B m_B}{48E_\gamma^2\,m_\rho^2\lambda_B} 
\e^{m_\rho^2/M^2}\,. 
\label{eq:xi-soft-t56}
\end{align}
Note that to our working accuracy one has to substitute $\phi_-(\omega)$ by 
the ``Wandzura-Wilczek contribution'' 
$\phi_-^{\WW}(\omega)$~\eqref{eq:WWmom}.

%
\section{Results}
\label{sec:N}
%

\begin{table}[t]
\renewcommand{\arraystretch}{1.2}
\begin{center}
    \begin{tabular}{| c | c | c | c |}
    \hline
    \hline
    $\mu_0$  &  1 GeV  &  & \\
    \hline
    $\Lambda^{(4)}_{\rm QCD}$ & $ 0.291552~\text{GeV}$ & $\alpha_s(\mu_0) $ & $ 0.348929$ \\
    \hline
    $\mu $ & $  (1.5 \pm 0.5)~\text{GeV}$ & $\mu_h $ & $ m_b/2 \div 2 m_b $ \\
    \hline
    $m_b $ & $ (4.8 \pm 0.1)~\text{GeV}$ & $\bar\Lambda$  &  $ m_B-m_b$\\
    \hline
    $\lambda_E^2/\lambda_H^2 $ & $ 0.5\pm 0.1 $ & $2\lambda_E^2+\lambda_H^2 $ & $ (0.25 \pm 0.15)~\text{GeV}^2$\\
    \hline
    $s_0$ & $(1.5 \pm 0.1)~\text{GeV}^2$& $M^2$ & $(1.25 \pm 0.25)~\text{GeV}^{2}$ \\
    \hline
    $\vev{\bar uu}(\mu_0)$ & $-(240\pm 15~\text{MeV})^3$ && \\ 
    \hline
    $m_B$ & 5.27929~GeV & $m_\rho$ & $0.77526$~GeV \\
    \hline
    $G_F$ & $1.166378\times10^{-5}~{\rm GeV}^{-2}$ & $\tau_B$ & $1.638\times10^{-12} s$ \\
    \hline
    $f_B $ & $ (192.0\pm 4.3)~\text{MeV}~\text{\cite{Aoki:2016frl}} $  & $|V_{ub}|^{\rm excl}$ & $(3.70\pm 0.16)\times10^{-3}~ \text{\cite{Patrignani:2016xqp}}$\\
    \hline
    \end{tabular}
\end{center}
\caption{Central values and ranges of all parameters used in this study.
The four-flavour $\Lambda_{\rm QCD}$ parameter corresponds to 
$\alpha_s(m_Z) = 0.1180$ with three-loop evolution and decoupling of the 
bottom quark at the scale $m_b$.}
\label{table:parameters}
\renewcommand{\arraystretch}{1.0}
\end{table}

In the numerical study presented below we use the NLL resummed result for 
the leading-power form factors~\cite{Beneke:2011nf} and the power-suppressed 
contributions $\xi\pm\Delta\xi$ in~\eqref{eq:FFs} given by the sum of 
hard-collinear higher-twist and soft corrections 
\begin{align}
\xi &= \xi^{\rm ht}\Big|_{\eqref{eq:HTcorrection}} 
+ \xi^{\rm soft}_{(\NLO)}\Big|_{\eqref{eq:xi-soft-NLO}} 
+ \xi^{\rm soft}_{(tw-3,4)}\Big|_{\eqref{eq:xi-soft-t34pre}} 
+ \xi^{\rm soft}_{(tw-5,6)}\Big|_{\eqref{eq:xi-soft-t56}}\,,
\notag\\
\Delta\xi &=\Delta\xi^{\rm ht}\Big|_{\eqref{eq:HTcorrection}} 
+\Delta \xi^{\rm soft}_{(tw-3,4)}\Big|_{\eqref{eq:xi-soft-t34bre}} 
+ \Delta \xi^{\rm soft}_{(tw-5,6)}\Big|_{\eqref{eq:xi-soft-t56}}\,. 
\label{eq:full-xi}
\end{align}
For the reader's convenience we have indicated the corresponding equation 
numbers. 

The nonperturbative inputs in the calculation have to be defined at a certain 
reference scale, $\mu_0$. As was done in previous work, we use $\mu_0 = 
1\,\text{GeV}$. Unless stated otherwise, the values of all scale-dependent 
hadronic parameters given below refer to this scale. In the calculation of 
the leading-power contributions to the form factors and the related soft 
correction $\xi^{\rm soft}_{(\NLO)}$ we evolve the inputs to the 
hard-collinear scale $\mu$, adopting $\mu=1.5~\text{GeV}$ as default. 
In the absence of the two-loop non-cusp anomalous dimension of the 
twist-2 $B$-meson LCDA $\phi_+(\omega)$, we perform the evolution in 
the LL approximation. Higher-twist contributions and the related soft 
corrections are always evaluated at the scale $\mu_0$.     
We use three-loop running of the strong coupling with $n_f=4$ active flavors. 
The central values and ranges of all parameters are collected in 
Table~\ref{table:parameters}. 

The principal input in our analysis is provided by the 
leading-twist $B$-meson LCDA $\phi_+(\omega)$. For the leading-power 
contribution to the form factors, the precise functional form of the LCDA 
is not important as it can be expressed in terms of the logarithmic 
moments\,\footnote{Note that our definition of the log-moments differs from 
those in \cite{Braun:2003wx} and \cite{Beneke:2011nf} 
by the substitution $\ln \mu/\omega \to \ln (e^{-\gamma_E}\lambda_B/\omega)$ 
and  $\ln \mu_0/\omega \to \ln (e^{-\gamma_E}\lambda_B/\omega)$, 
respectively. The purpose of this change is to decorrelate the log-moments 
from the value of $\lambda_B$ in the models for $\phi_+(\omega)$ considered 
below.} 
\begin{align}
\widehat{\sigma}_n = \int^\infty_0 d\omega\,\frac{\lambda_B}{\omega}\,
\ln^n\frac{\lambda_B e^{-\gamma_E}}\omega\,\phi_+(\omega)
\label{sigma-n}
\end{align}
with $\sigma_0 =1$ defining $\lambda_B$. To the NLL accuracy only the values 
of $\lambda_B$, $\sigma_1$ and $\sigma_2$ are needed. The LCDA and the 
moments are renormalization scale dependent. For brevity, we omit the 
explicit scale argument $\mu$.

In contrast, the evaluation of the soft (endpoint) contributions requires 
the full functional form of the LCDAs. We will use three two-parameter 
families of functions to assess the model dependence of the soft 
contribution. In the $s$-space 
representation \eqref{Phi+Phi-}~\cite{Bell:2013tfa,Braun:2014owa}
\begin{subequations}
\begin{align}
\eta_+^{(\mI)}(s)&= {}_1F_1(1+2/b,2/b,-s\omega_0) =  
\big(1-\frac12 b s\omega_0\big) e^{-s\omega_0}\,,\qquad  0\leq b\leq1\,,
\label{etaI}\\[0.2cm]
\eta_+^{(\mII)}(s)&=  {}_1F_1(2+a,2,-s\omega_0)\,, \qquad -0.5<a<1\,,
\label{etaII}\\[0.25cm]
\eta_+^{(\mIII)}(s)&=
{}_1F_1(3/2+a,3/2,-s\omega_0)\,, \qquad  0<a<0.5\,, 
\label{etaIII}
\end{align}
\label{eta+models}
\end{subequations}
corresponding in momentum space to
\begin{subequations}
\label{phi+models}
\begin{align}
\phi_+^{(\mI)}(\omega)&=
\left[(1-b)+\frac{b\,\omega}{2\omega_0}\right]\frac{\omega}{\omega_0^2}\,\e^{-\omega/\omega_0}\,, 
\label{modelI}\\
\phi_+^{(\mII)}(\omega)&
=\frac{1}{\Gamma(2+a)}\frac{\omega^{1+a}}{\omega_0^{2+a}}\e^{-\omega/\omega_0}\,, 
\label{modelII}\\
\phi_+^{(\mIII)}(\omega)&
=  \frac{\sqrt{\pi}}{2 \Gamma(3/2+a)} \frac{\omega}{\omega_0^2}\,e^{-\omega/\omega_0} 
 \,U(-a,3/2-a,\omega/\omega_0)\,,
\label{modelIII}
\end{align}
\end{subequations}
where ${}_1F_1(\alpha,\beta,z)$ is a hypergeometric function, and 
$U(\alpha,\beta,z)$ the confluent hypergeometric function of the second kind. 
The above functional forms are assumed to hold at $\mu_0=1\,$GeV.

\begin{figure}[t]
\centering
 \includegraphics[width=.99\linewidth]{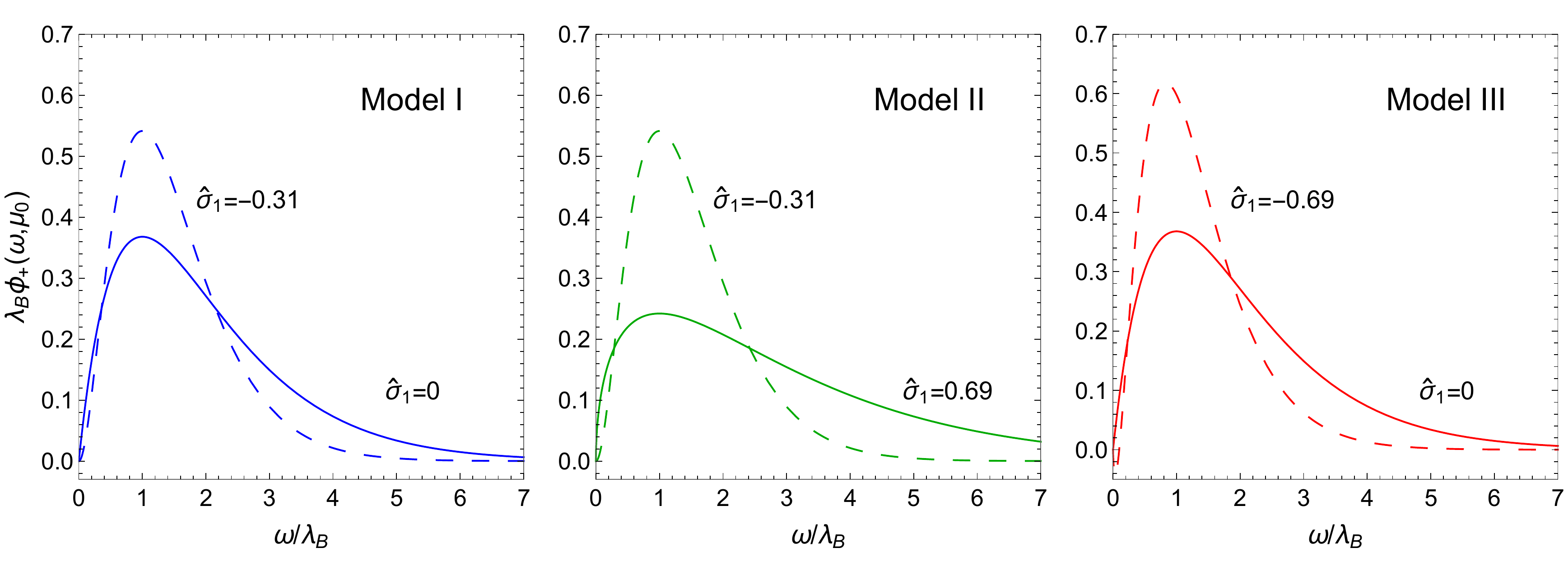}
\caption{\small 
$B$-meson leading-twist LCDA $\lambda_B \phi_+(\omega,\mu_0)$ 
for the three models described in the text.
}
\label{fig:models}
\end{figure} 

The three models in~\eqref{phi+models} for the limiting values of the 
parameters \eqref{eta+models} are shown in Fig.~\ref{fig:models}. They
can be viewed as particular cases of the more general three-parameter ansatz 
\begin{align}
\label{Gmodel}
\eta_+(s) &= {}_1F_1(\alpha,\beta,-s\omega_0)\,,\qquad \alpha,\beta > 1\,,
\notag\\[0.2cm]
  \phi_+(\omega) &=  \frac{\Gamma(\beta)}{\Gamma(\alpha)} \frac{\omega}{\omega_0^2}\,e^{-\omega/\omega_0} 
 U(\beta-\alpha,3-\alpha,\omega/\omega_0)\,.
\end{align}
For this ansatz 
\begin{align}
\lambda_B &= \frac{\alpha-1}{\beta-1} \omega_0\,, \qquad 
\widehat \sigma_1 = \psi(\beta-1) - \psi(\alpha-1) + 
\ln \frac{\alpha-1}{\beta-1}\,,  
\end{align}
etc., so that the only dimensionful parameter $\omega_0$ can be traded for 
$\lambda_B$ and the logarithmic moments defined in \eqref{sigma-n} depend on 
the ``shape parameters'', $\alpha$ and $\beta$.
  
The particular choices \eqref{eta+models}, \eqref{phi+models} are motivated 
by the experience in the modelling of the pion LCDA, where especially the 
endpoint behaviour came under scrutiny in connection with the BaBar and 
BELLE measurements of the $\gamma^*\to\pi\gamma$ transition form factor, 
see e.g.~\cite{Agaev:2010aq,Agaev:2012tm,Cloet:2013tta,Stefanis:2015qha}. 
The parameter range indicated in~\eqref{eta+models} corresponds to 
$- 0.306853 < \widehat{\sigma}_1 < 0$ for Model~I, $- 0.306853 < \widehat{\sigma}_1 < 0.693147$ for Model~II and 
$- 0.693147 < \widehat{\sigma}_1 < 0$ for Model~III, so that,   
taken together, they cover the range
\begin{align}
   - 0.693147 < \widehat{\sigma}_1 < 0.693147
\label{sigma1range}
\end{align}
for arbitrary $\lambda_B$. The value $\widehat{\sigma}_1=0$ 
corresponds to the simple exponential model 
$\phi_+(\omega) = (\omega/\omega_0^2) \,e^{-\omega/\omega_0}$ suggested 
in~\cite{Grozin:1996pq}. 

The large-momentum behaviour of the $B$-meson LCDA can be studied in 
perturbation theory in a cutoff scheme~\cite{Lee:2005gza}. In this way the 
first moment $\int_0^{\mu_{F}}d\omega\,\omega\phi_+(\omega)$ is related to a 
properly defined $\bar\Lambda(\mu_{F}) =m_B-m_b(\mu_{F})$ and  
the second moment, $\int_0^{\mu_{F}}d\omega\,\omega^2\phi_+(\omega)$, 
to matrix elements of the quark-gluon operators~\eqref{def:lambdaEH}, which 
were estimated with QCD sum rules~\cite{Grozin:1996pq,Nishikawa:2011qk}.
However, it was shown in~\cite{Feldmann:2014ika} that such relations do not 
generally provide significant constraints on the logarithmic moments 
$\widehat\sigma_1$, $\widehat\sigma_2$, since they can be satisfied by 
adding a large-momentum ``tail'' to any given (reasonable) model for 
$\phi_+(\omega)$. Following this argument, we will assume that the 
``true'' LCDA can be written as
\begin{align}
  \phi_+(\omega,\mu) = \phi^{\rm\scriptscriptstyle model}_+(\omega,\mu) 
+ \delta\phi_+(\omega,\mu)\,, 
\end{align}
where $\phi^{\rm\scriptscriptstyle model}_+(\omega,\mu)$ refers to one of the 
models specified in \eqref{phi+models} and the added ``tail'' is 
concentrated at large momenta $\omega \gg \lambda_B$. Its role is to 
ensure that the relations for the first two moments are satisfied to the 
required accuracy. We assume that this additional term  can be chosen in 
such a way that the first few logarithmic moments are not 
affected~\cite{Feldmann:2014ika}. In this case an explicit expression for 
$\delta\phi_+(\omega,\mu)$ is not needed as it does not enter any of the 
three contributions to the form factors: neither (1) the perturbative 
leading-twist leading-power contribution, as it is expressed in terms of the 
logarithmic moments, nor (2) the soft corrections, as they originate from 
small momenta, nor (3) higher-twist corrections, as they are expressed 
directly in terms of $\bar\Lambda$ and higher-twist matrix  elements 
$\lambda_E^2$, $\lambda_H^2$ (see below).\footnote{We must assume that 
$\delta\phi_+$ decreases sufficiently fast
at $\omega\to\infty$ so that its first few moments are finite.
While this cannot hold true in general due to
perturbative radiative corrections~\cite{Lange:2003ff},
the assumption is necessary for consistency of tree-level
calculations of higher-twist contributions as performed here. See also the 
Appendix.}

In Ref.~\cite{Braun:2017liq} several models for the higher-twist LCDAs have 
been suggested that have the expected low-momentum behaviour and satisfy the 
(tree-level) EOM constraints. One can show that these models can be obtained 
as particular cases of the more general ansatz~\eqref{T4ansatz}. For 
these models one obtains the remarkably simple expression
\begin{align}
     \xi^{\rm ht} (E_\gamma)  &=
  - \frac{e_u f_B m_B }{2 E^2_\gamma}\biggl\{ \frac{2(\lambda_E^2+ 2\lambda_H^2)}{6\bar\Lambda^2+ 2\lambda_E^2+ \lambda_H^2}+ \frac12 \biggr\}
\notag\\&\quad+
\frac{e_u f_B m_B}{4 m_bE_\gamma}\biggl\{\frac{\bar\Lambda }{\lambda_B}  -2 
+ \frac{4(\lambda_E^2-\lambda_H^2)}{6\bar\Lambda^2+ 2\lambda_E^2+ \lambda_H^2}
      \biggr\}, 
    \label{eq:HTmodel}
\end{align}
and the higher-twist correction does not depend on the functional form of 
the profile function $f(\omega)$ in the ansatz~\eqref{T4ansatz}. 

We use the range $m_b = 4.7 \div 4.9\,\text{GeV}$ for the pole mass and 
define $\bar\Lambda = m_B - m_b$. It has to be mentioned that the derivation 
of the higher-twist corrections is based on equations of motion at tree 
level~\cite{Kawamura:2001jm,Braun:2017liq}. For consistency, the relations 
between moments of the LCDA and local matrix elements have to be assumed at 
tree level as well, Eq.~\eqref{GN}. The scheme-dependence of $\bar\Lambda$ 
and our result  \eqref{eq:HTcorrection}, \eqref{eq:HTmodel} for the 
higher-twist correction should be cancelled by a correction proportional to 
$\mu_F \alpha_s/\pi$ that has not been calculated so far. Before this is 
done, the numerical value of $\bar\Lambda$ (or, equivalently, of the 
$b$-quark pole mass $m_b$) should be viewed as an educated guess.

The matrix elements $\lambda_E^2$ and  $\lambda_H^2$ are defined
in~\eqref{def:lambdaEH}. The existing QCD sum rule 
estimates~\eqref{QCDSR:lambdaEH} fall in the range 
\begin{align}
 0.1~\text{GeV}^2 < 2\lambda_E^2+\lambda_H^2 < 0.4~\text{GeV}^2,
&&
 \lambda_E^2/ \lambda^2_H = 0.5\pm 0.1\,. 
\end{align}
For this range of values, the dependence of the higher-twist correction 
in~\eqref{eq:HTmodel} on $\lambda_E^2$ and $\lambda_H^2$ is rather weak so 
that a large uncertainty in the matrix elements does not play a major role, 
except for large $\lambda_B$.

In order to understand the qualitative features of soft corrections let us 
consider the leading-order twist-two contribution 
$\xi^{\rm soft}_{(\LO)}$~\eqref{eq:xi-soft-LO} as an example. Normalizing to 
the leading-order QCD result~\eqref{eq:FFs}, and extracting the expected 
$1/(2E_\gamma)$ suppression factor we define~\cite{Braun:2012kp}
\begin{align}
 \xi^{\rm soft}_{(\LO)}(E_\gamma) &= 
\frac{e_u f_B m_B}{2 E_\gamma \lambda_B(\mu)} \,U_{\rm LL}\,
\frac{\widehat\xi^{\rm soft}_{(\LO)}(E_\gamma)}{2 E_\gamma}\,,
\notag\\
\widehat\xi^{\rm soft}_{(\LO)}(E_\gamma) &= 
2 E_\gamma \lambda_B(\mu)\int\limits_0^{s_0/2E_\gamma}\!d\omega
\left[ \frac{2 E_\gamma}{m_\rho^2} e^{-(2E_\gamma \omega - m^2_\rho)/M^2} 
- \frac{1}{\omega}\right] \phi_+(\omega,\mu)\,.
\label{eq:xi-tilde}
\end{align}
This expression involves two parameters --- the continuum threshold $s_0$ 
and the Borel parameter $M^2$ --- which we choose in the range
\begin{align}
 1.4\,\text{GeV}^2 < s_0 < 1.6\,\text{GeV}^2, \qquad  
 1.0\,\text{GeV}^2 < M^2 < 1.5\,\text{GeV}^2. 
\end{align} 
The soft correction originating from twist-five and twist-six LCDAs depends
 in addition on the quark condensate
$\langle \bar u u\rangle (1~\text{GeV}) = - (240\pm 15~\text{MeV})^3$. 

In the asymptotic regime $\mu^2 \sim \Lambda_{\rm QCD} E_\gamma \to \infty$ 
the LCDA $\phi_+(\omega,\mu)$ is driven by the renormalization group flow to 
linear behaviour $\phi_+(\omega) \sim \omega \phi_+'(0)$ for 
$\omega\to 0$ independent on the initial condition at low scales. In this 
case $\widehat\xi^{\rm soft}_{(\LO)}(E_\gamma) \to \,\text{const}\cdot 
\lambda_B \phi'_+(0) + \mathcal{O}(1/E_\gamma)$ so that the soft correction 
is proportional to the (in this limit finite) derivative of the LCDA at zero 
momentum. For physically interesting photon momenta 
$E_\gamma\sim 1.5-2.5$~GeV the dominance of the $\omega\to0$  region does 
not hold since the integration in \eqref{eq:xi-tilde} goes over the 
momentum region $0 < \omega < 300\div 500\,\text{MeV}$ that is comparable 
with the characteristic momentum scale $\lambda_B$ in the LCDA. 
Thus the integral is determined by global properties of the LCDA 
(normalization, width, etc.) rather than the endpoint behaviour. The 
situation is similar in this respect to the better studied 
reaction~$\gamma^*\gamma\to\pi$ in which case it was 
shown~\cite{Agaev:2010aq} that an anomalous endpoint behaviour of the 
pion LCDA cannot explain by itself the strong scaling violation observed by 
BaBar~\cite{Aubert:2009mc} up to much higher scales 
$Q^2 \sim 20\div 30\,\text{GeV}^2$.  

\begin{figure}[t]
\centering
 \includegraphics[width=.99\linewidth]{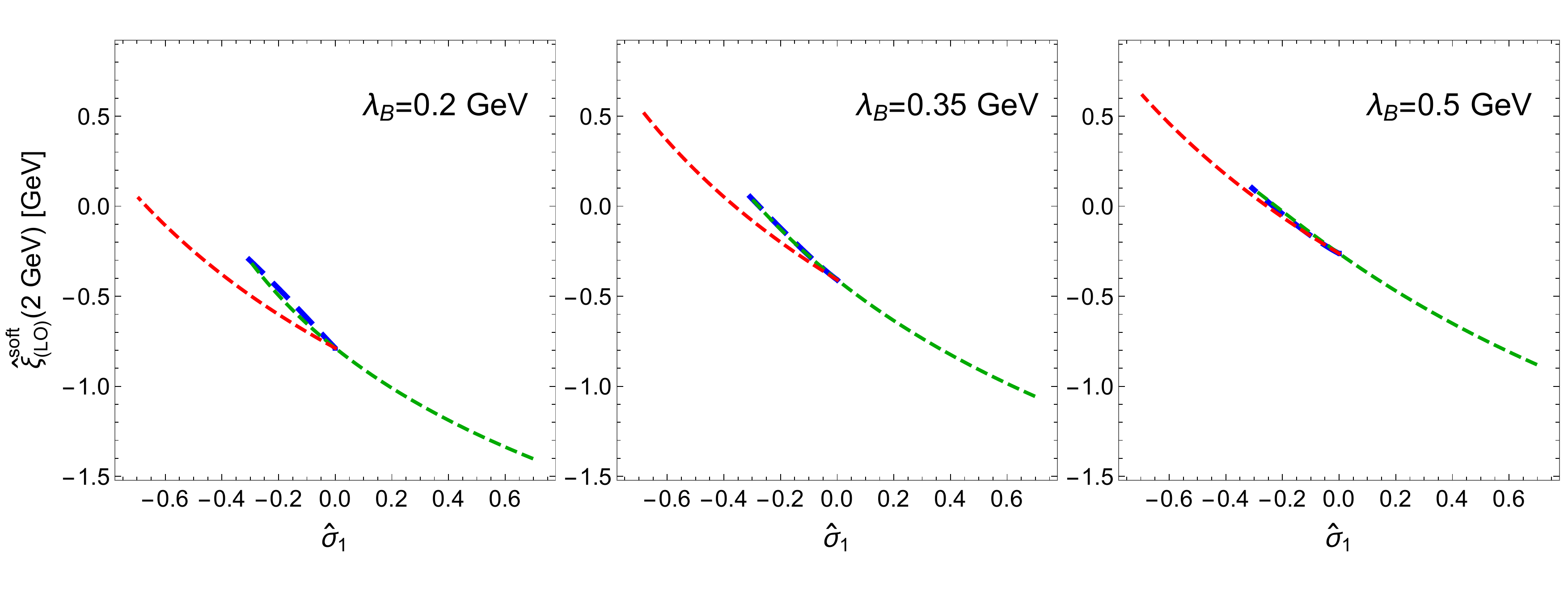}
\caption{\small 
The leading-order soft correction normalized to the corresponding QCD result,
$\widehat\xi^{\rm soft}_{(\LO)}(2\,\text{GeV})$ \eqref{eq:xi-tilde}, as a 
function of the first logarithmic moment $\widehat\sigma_1$ \eqref{sigma-n} 
for the three models of the leading-twist B-meson LCDA defined in 
\eqref{modelI} (blue), \eqref{modelII} (green) and \eqref{modelIII} (red), 
respectively, and for three different values of $\lambda_B$ as specified 
on the plots.}
\label{fig:xi0}
\end{figure} 

As already noticed in~\cite{Braun:2012kp},\, the normalized soft correction 
$\widehat\xi^{\rm soft}_{(\LO)}(E_\gamma)$ \eqref{eq:xi-tilde} 
depends only weakly on photon energy $E_\gamma$ (in the relevant range). For 
illustration we plot 
$\widehat\xi^{\rm soft}_{(\LO)}(E_\gamma = 2\,\text{GeV})$ in 
Fig.~\ref{fig:xi0} as a function of $\widehat\sigma_1$ for three different 
values of $\lambda_B$ and central values  of the sum rule parameters, 
$s_0=1.5\,\text{GeV}^2$ and $M^2=1.25\,\text{GeV}^2$. The blue, green and 
red curves are obtained using models I, I\!I, and I\!I\!I 
in~\eqref{phi+models}, respectively, with the indicated parameter range.
Note that this correction can be  quite sizable, e.g., the value 
$\widehat\xi^{\rm soft}_{(\LO)} = -1.0\,\text{GeV}$ 
corresponds to a power-suppressed contribution to the form factors of the 
order of $ (-1.0\,\text{GeV})/(2E_\gamma)$ with respect to the leading-order,  
leading-twist result. It attracts attention that 
$\widehat\xi^{\rm soft}_{(\LO)}(2\,\text{GeV})$ can be both positive and 
negative, and depends strongly on the value of the first logarithmic moment, 
$\widehat\sigma_1$. For a given $\widehat\sigma_1$, the correction is fairly 
close in all three models (in the regions where there is an overlap).
This agreement is trivial for $\widehat\sigma_1=0$ as all models reduce 
to the same simple exponential model, but it is not trivial for the whole 
range. Note also that the precise small-$\omega$ behaviour of the LCDA is 
irrelevant: for model I\!I the derivative $\phi'_+(0)$ is 
changing from zero to infinity as the parameter $a$ 
and $\widehat\sigma_1$ change sign, with no visible effect on the result.

\begin{figure}[t]
\centering
  \includegraphics[width=.99\linewidth]{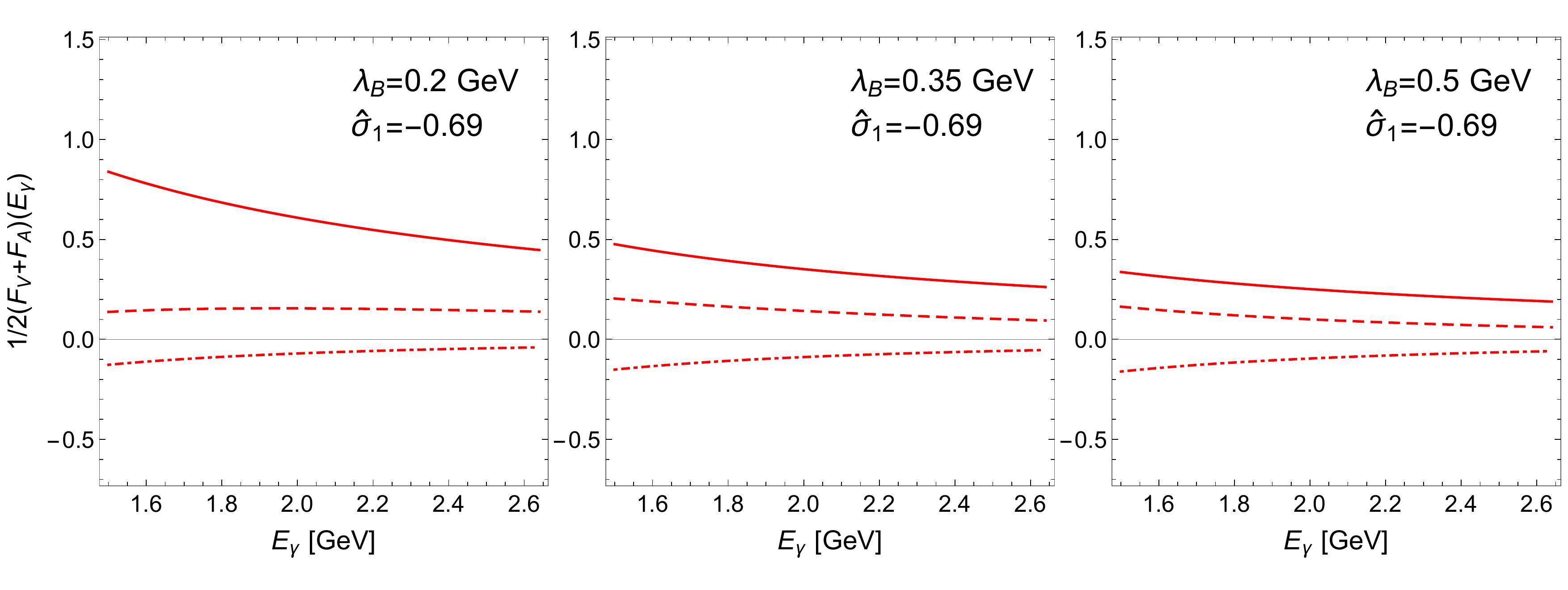}\\[-4mm]
  \includegraphics[width=.99\linewidth]{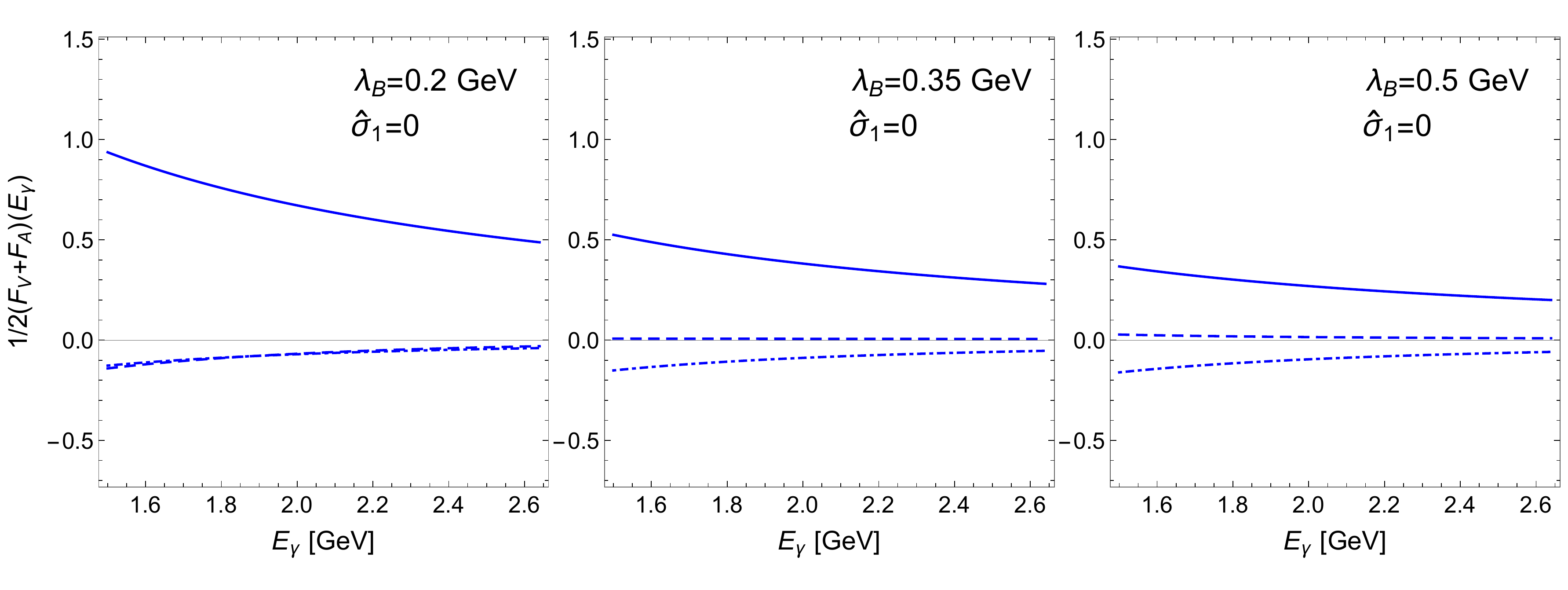}\\[-4mm]
  \includegraphics[width=.99\linewidth]{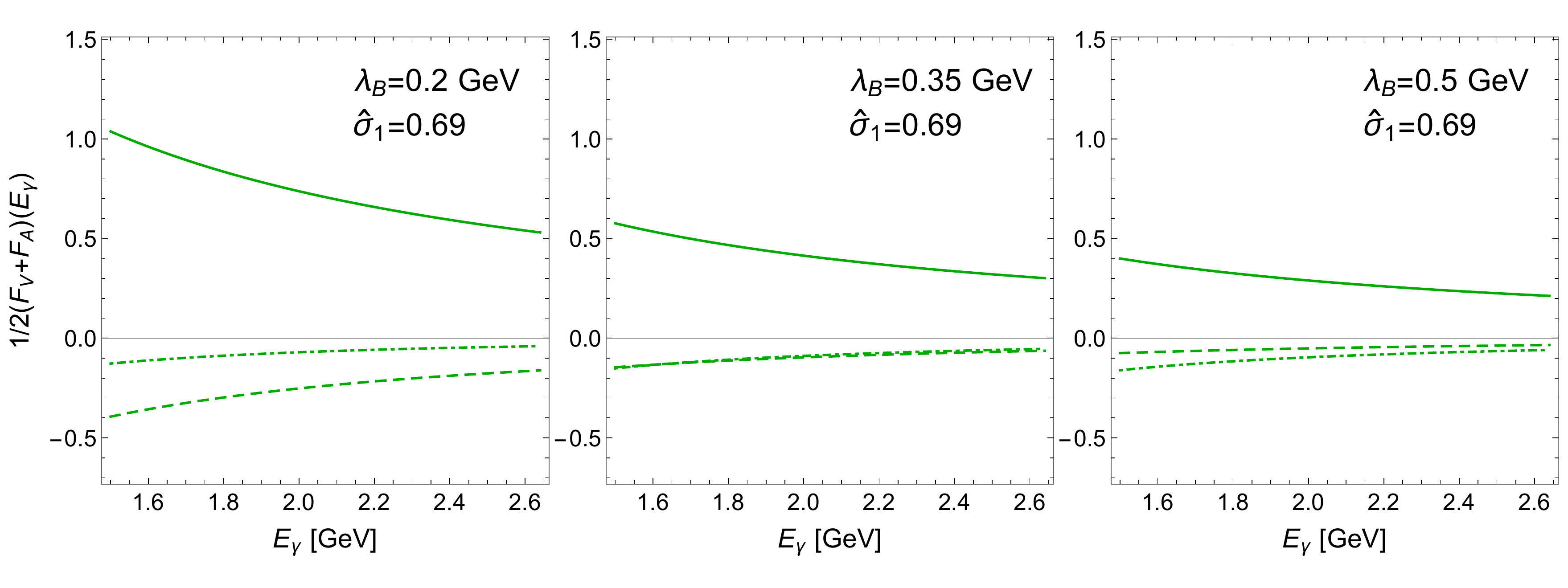}
\caption{\small Perturbative (solid curves), soft (dashed curves)
and higher-twist (dash-dotted curves) contributions to the form factor 
$(1/2)(F_V+F_A)$ as functions of photon energy $E_\gamma$ for different 
choices of the parameters $\lambda_B$ and $\sigma_1$.
The colour coding corresponds to Fig.~\ref{fig:models}.
}
\label{fig:pert-Ht-soft}
\end{figure} 

The relative size of various contributions to the ``symmetry-preserving'' 
form factor combination $(F_V+F_A)/2$ (alias the helicity form factor $F_-$) 
is illustrated in Fig.~\ref{fig:pert-Ht-soft} for several choices of the 
parameters $\lambda_B$ and $\sigma_1$. We show the NLL resummed perturbative 
result~\cite{Beneke:2011nf} (solid curves), the total soft correction 
$\xi^{\rm soft} = \xi^{\rm soft}_{(\NLO)}+\xi^{\rm soft}_{(tw-3,4)}
+\xi^{\rm soft}_{(tw-5,6)}$,  and the hard-collinear higher-twist 
correction $\xi^{\rm ht}$ by the solid, dashed and dash-dotted curves, 
respectively. One sees that the higher-twist correction is negative and 
relatively small for all cases, whereas the soft correction can be of either 
sign and for small $\lambda_B$ becomes rather large. The effect of the soft 
correction is always to counteract the change of the perturbative
contribution due to the variation of $\lambda_B$ and, in particular, 
$\sigma_1$ so that the sensitivity of the form factor to the model of the 
LCDA is reduced upon accounting for the soft correction as compared 
to the leading-power result alone. Among the different contributions to the 
soft correction the part related to the leading-twist LCDA, 
$\xi^{\rm soft}_{(\NLO)}$~\eqref{eq:xi-soft-NLO}, is dominant in all cases;
the other two contributions are relatively small. In particular 
$\xi^{\rm soft}_{(tw-5,6)}$ is at most 6\% of the total value. This is 
reassuring, suggesting that soft contributions related to the LCDAs of 
even higher twist can be small as well, and also because the approximation 
leading to {\eqref{eq:xi-soft-t56}} is rather crude.  
We also find that the $1/m_b$ power corrections are 
generally much smaller than the $1/E_\gamma$ corrections, and so are the 
``genuine'' three-particle higher-twist corrections relative to those 
that can be related to two-particle terms by the equations of motion.   

\begin{figure}[t]
\centering
  \includegraphics[width=.99\linewidth]{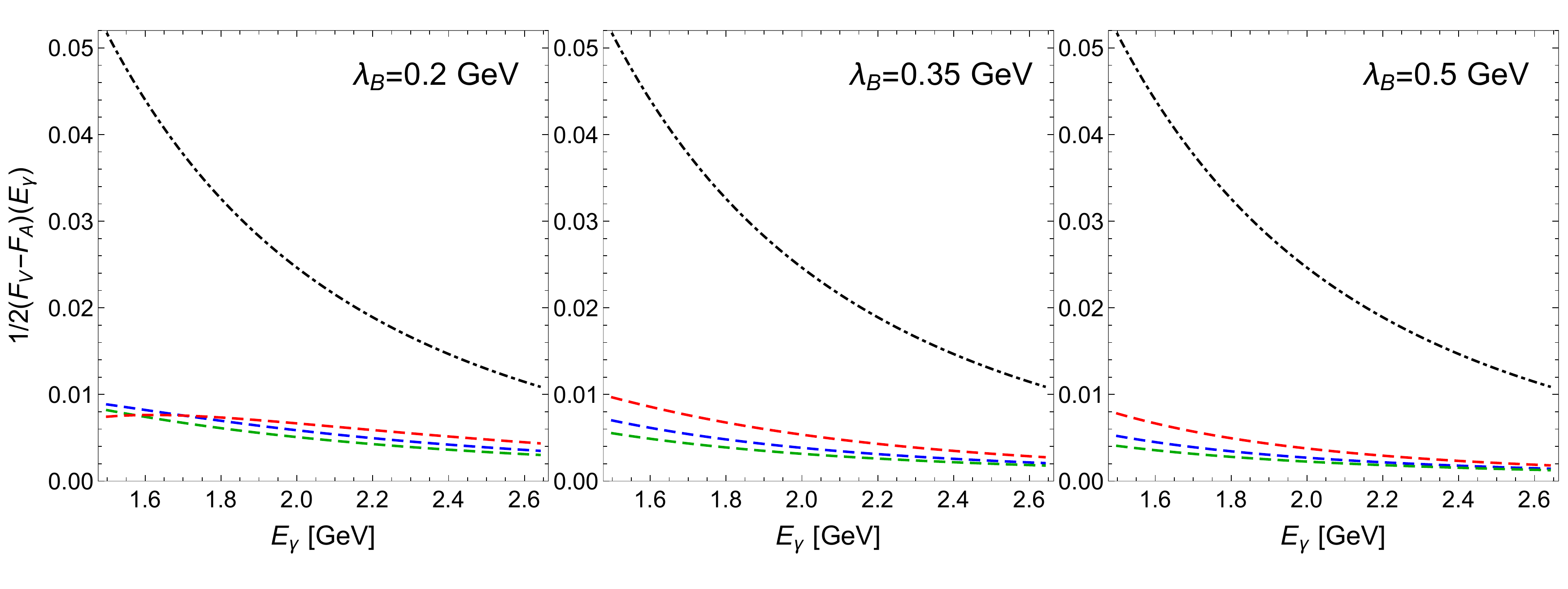}
\caption{\small Model-independent  higher-twist \text{\cite{Beneke:2011nf}} 
(black dash-dotted curves) and soft (dashed curves) contributions to the 
form factor difference $(1/2)(F_V-F_A)$ for three choices of $\lambda_B$.
The soft corrections are shown  for $\widehat\sigma_1 = -0.69, 0, +0.69$ 
in red, green and blue colour, respectively. The colour coding corresponds 
to Fig.~\ref{fig:models}.
}
\label{fig:Deltaxi}
\end{figure} 

A similar decomposition of the various contributions to the 
``symmetry-breaking'' form factor difference $\Delta\xi = (F_V-F_A)/2$ is 
shown in Fig.~\ref{fig:Deltaxi}. It is dominated by the model-independent 
higher-twist correction \eqref{eq:symbreleading}~\cite{Beneke:2011nf}
(black dash-dotted curves) whereas the soft contributions (dashed) turn out 
to be small in all cases. They are shown in three colours corresponding to 
the choice $\widehat\sigma_1 = -0.69, 0, +0.69$ at the boundaries and in 
the middle of the three models' envelope.

To visualize the relative importance of different uncertainties due to the 
choice of the parameters in the range specified in 
Table~\ref{table:parameters}, we consider the vector form factor $F_V$ for 
$E_\gamma=2$~GeV, and $\lambda_B=0.35$~GeV, in the middle of the range of 
interest, and two extreme values for the first logarithmic 
moment, $\widehat\sigma_1 = \pm 0.693$. We obtain
\begin{align}
& F_V(E_\gamma=2\,\text{GeV}, \lambda_B=0.35\,\text{GeV}, 
\widehat\sigma_1 = 0.693) 
\notag\\ &\hspace*{0.5cm}=\, 
0.258\,+\,
\begin{pmatrix}+0.012 \\-0.017 \end{pmatrix}_{m_b} 
\!\!+\begin{pmatrix}+0.000 \\-0.007 \end{pmatrix}_{\mu} 
\!\!+\begin{pmatrix}+0.006 \\-0.006 \end{pmatrix}_{\mu_h}
\!\!+\begin{pmatrix}+0.001 \\ -0.000\end{pmatrix}_{M^2}
\!\!+\begin{pmatrix}+0.001 \\ -0.001\end{pmatrix}_{s_0}
\notag\\&\hspace*{2.25cm}
+\begin{pmatrix}+0.016 \\ -0.013\end{pmatrix}_{2\lambda_E^2+\lambda_H^2}
\!\!\!\!+\begin{pmatrix}+0.002 \\ -0.003\end{pmatrix}_{\lambda_E^2/\lambda_H^2}
\!\!\!\!+\begin{pmatrix}+0.004\\ -0.003\end{pmatrix}_{\langle \bar u u\rangle}  
\,=\,0.258^{+0.021}_{-0.024}\,, 
%
\notag\\[0.3cm]
 & F_V(E_\gamma=2\,\text{GeV}, \lambda_B=0.35\,\text{GeV}, \widehat\sigma_1 = -0.693)
 \notag\\ &\hspace*{0.5cm}=\, 
0.435\,+\,
\begin{pmatrix}+0.013 \\-0.017\end{pmatrix}_{m_b} 
\!\!+\begin{pmatrix}+0.000 \\-0.006\end{pmatrix}_{\mu} 
\!\!+\begin{pmatrix}+0.010 \\-0.009\end{pmatrix}_{\mu_h}
\!\!+\begin{pmatrix}+0.013 \\ -0.018\end{pmatrix}_{M^2}
\!\!+\begin{pmatrix}+0.003 \\ -0.004\end{pmatrix}_{s_0}
\notag\\&\hspace*{2.25cm}
+\begin{pmatrix}+0.014 \\ -0.011\end{pmatrix}_{2\lambda_E^2+\lambda_H^2}
\!\!\!\!+\begin{pmatrix}+0.002 \\ -0.002\end{pmatrix}_{\lambda_E^2/\lambda_H^2}
\!\!\!\!+\begin{pmatrix}+0.004\\ -0.003\end{pmatrix}_{\langle \bar u u\rangle}  
\,=\,0.435^{+0.025}_{-0.030}\,,
\end{align}
%
%
where we added the errors in quadrature to arrive at the final numbers.
We do not include here the uncertainty due to the $B$-meson decay constant
$f_B$, cf.~Table~\ref{table:parameters}, which enters as an overall factor, 
and can therefore trivially be added. Apart from this, the overall 
uncertainty is only about 6-9\%, with the main contributions 
from the $b$-quark mass (alias $\bar\Lambda$), the Borel parameter, 
and the twist-four matrix element $2\lambda_E^2+\lambda_H^2$. 
The hard-collinear factorization scale ($\mu$) dependence of 
$\xi^{\rm soft}_{(\NLO)}$ turns out to be large 
(up to 30\%) but is always anticorrelated with the scale dependence of the 
leading-power contribution such that the $\mu$-dependence of their 
sum is reduced compared to that of the leading power term alone.    

\begin{figure}[t]
\centering
  \includegraphics[width=.99\linewidth]{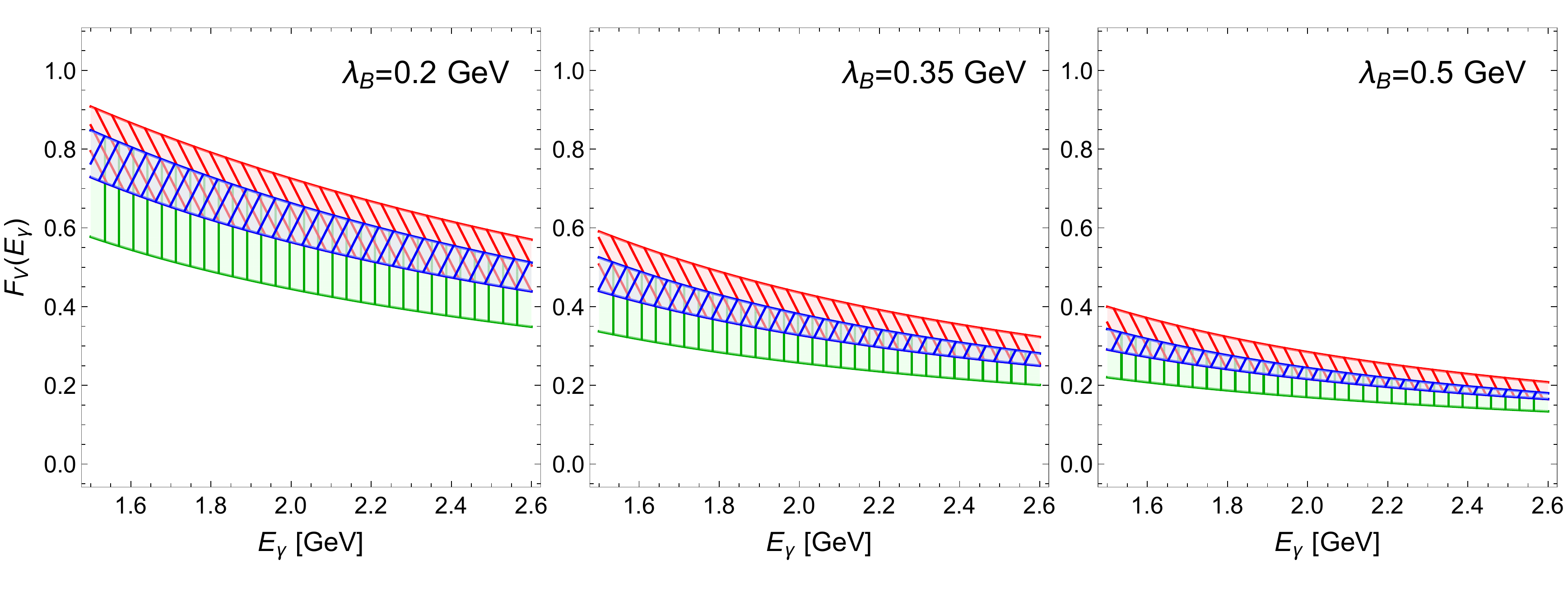}\\[-4mm]
  \includegraphics[width=.99\linewidth]{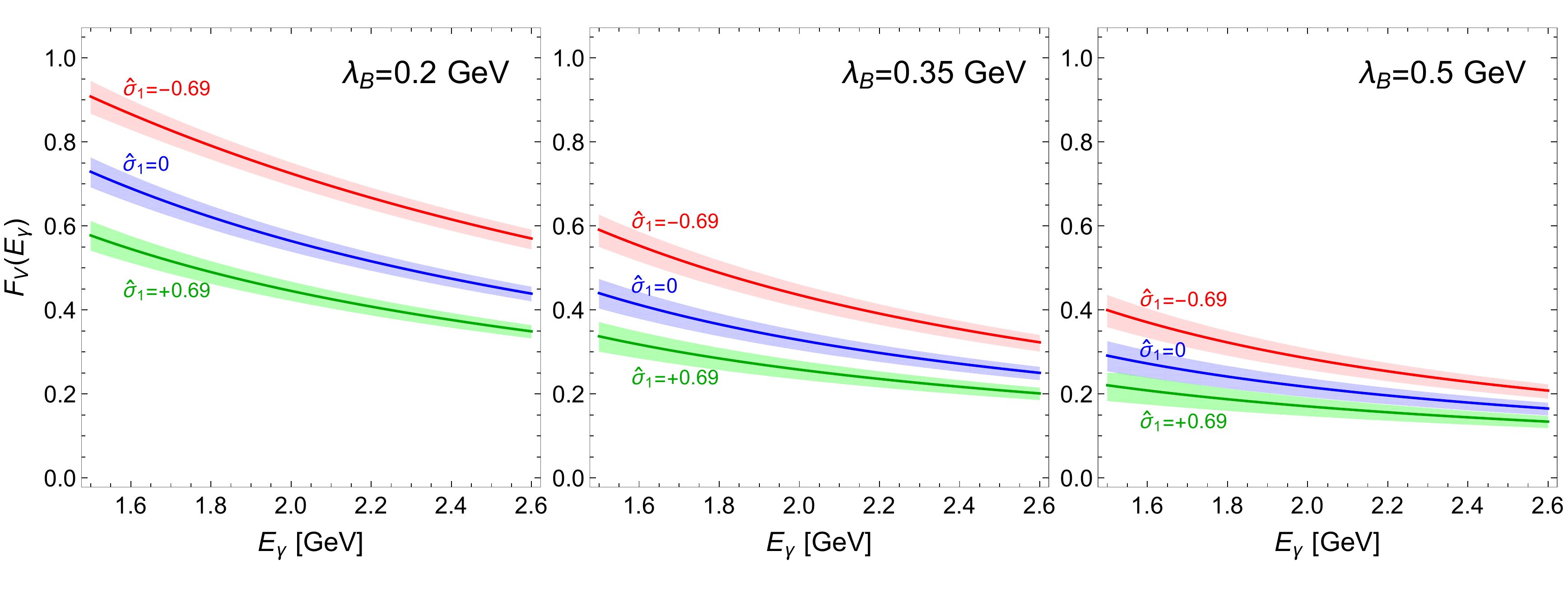}
\caption{\small Vector form factor $F_V(E_\gamma)$. 
The shaded regions on the upper panels show the variation for a given model 
with the range of parameters specified in \eqref{eta+models}. The uncertainty 
due to other parameters in the range specified in 
Table~\ref{table:parameters} is shown on the three lower panels for 
$\widehat\sigma_1 = \pm 0.69$ corresponding to the boundary of the models' 
envelope in the upper plot, and for $\widehat\sigma_1 = 0$ corresponding 
to the simple exponential ansatz~\cite{Grozin:1996pq}.
The colour coding corresponds to Fig.~\ref{fig:models}.
}
\label{fig:FV}
\end{figure} 
\begin{figure}[t]
\centering
  \includegraphics[width=.99\linewidth]{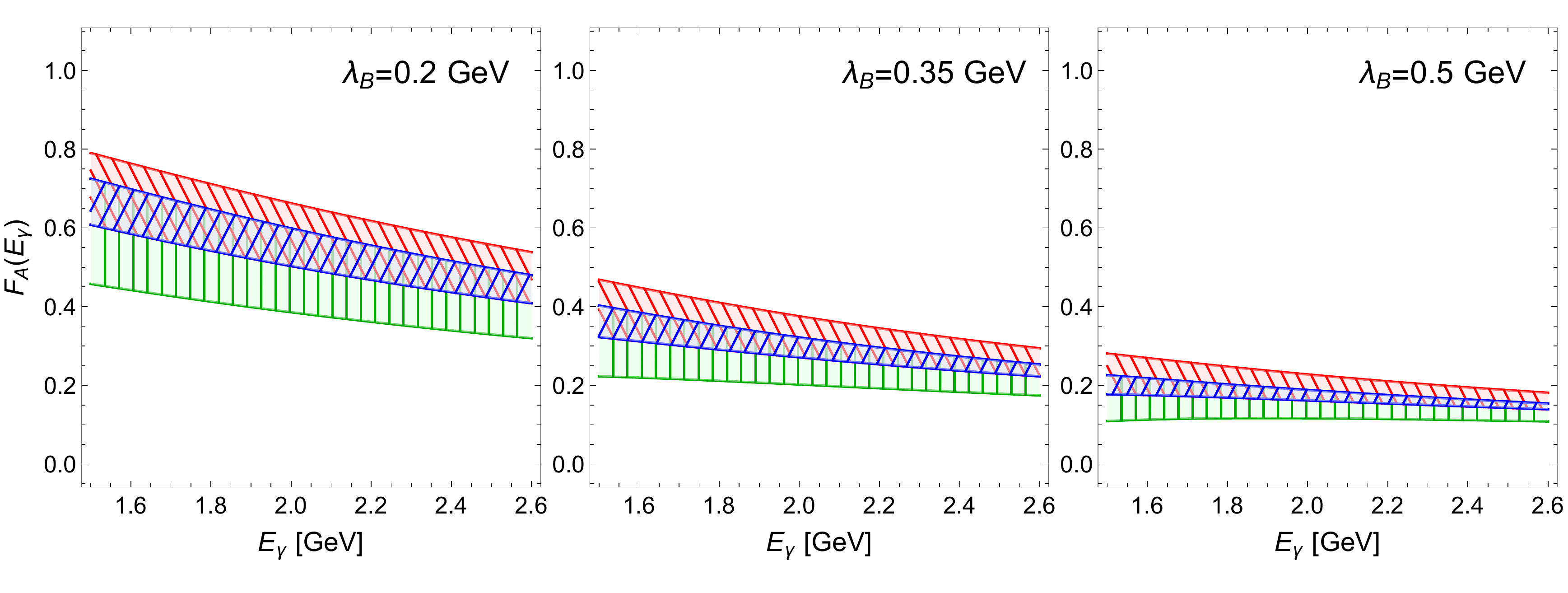}\\[-4mm]
  \includegraphics[width=.99\linewidth]{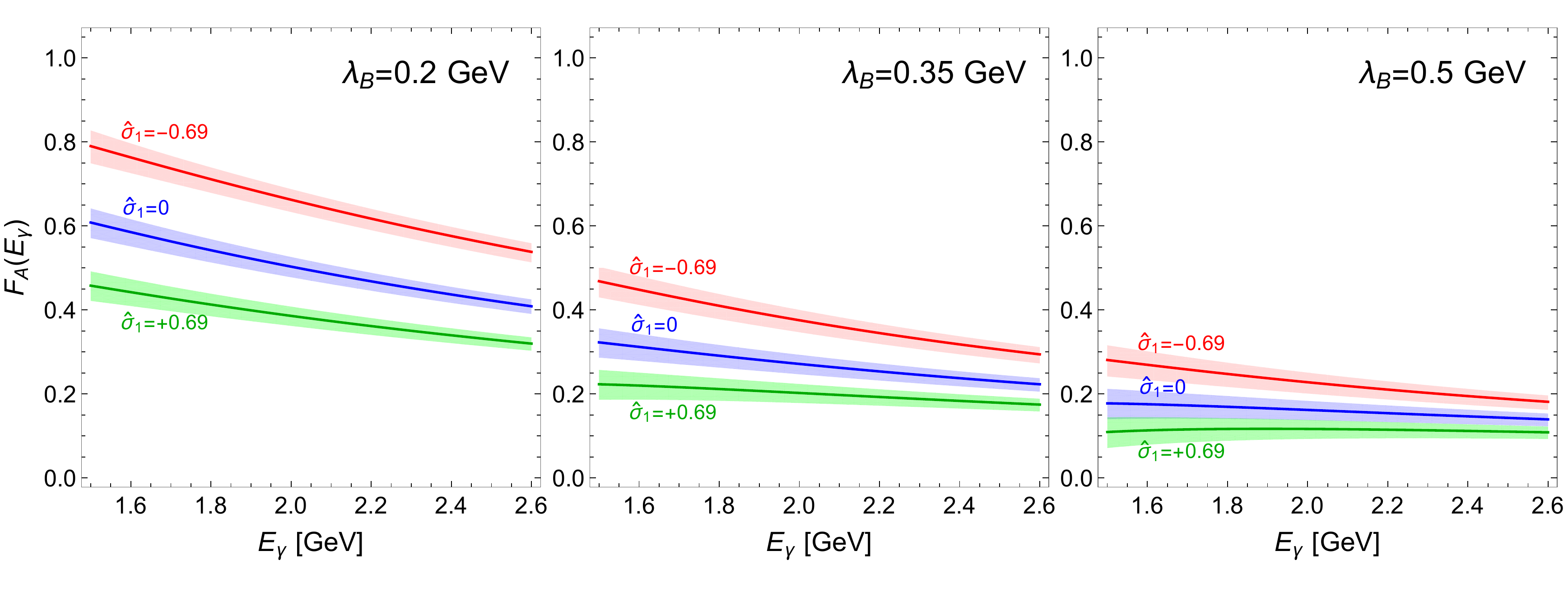}
\caption{\small Axial form factor $F_A(E_\gamma)$. The legend follows 
Fig.~\ref{fig:FV}. 
}
\label{fig:FA}
\end{figure} 

Our final results for the vector $F_V$ and axial $F_A$ form factors are 
shown in Figs.~\ref{fig:FV} and~\ref{fig:FA}, respectively.\footnote{We 
recall that the contribution of photon emission from the final state lepton 
is not included in $F_A$, cf.~\eqref{eq:width}.} 
The shaded regions on the upper panels in both figures show the variation 
for a given model with the range of parameters specified in \eqref{eta+models}
and central values for other parameters. The colour coding follows 
Fig.~\ref{fig:models}. The uncertainty from variation of the other 
parameters in the range specified in Table~\ref{table:parameters} is shown 
on the lower panels for three cases: $\widehat\sigma_1 = \pm 0.69$ 
corresponding to the boundaries of the three models' envelope,
and $\widehat\sigma_1=0$. For the last value our three models coincide and
reduce to the simple exponential model of Ref.~\cite{Grozin:1996pq}. 
This uncertainty is below 15\% in all cases.

Two important conclusions can be drawn from these results. First, the 
uncertainty from all parameters except those of the leading-twist 
$B$-meson LCDA $\phi_+(\omega)$ is generally smaller than the dependence 
on $\phi_+(\omega)$ itself, which is large. This is welcome, since the 
measurement of the $B\to \gamma \ell \nu_\ell$ process is primarily 
seen as a means to determine the $B$-meson LCDA $\phi_+(\omega)$, 
in particular $\lambda_B$. The calculation of the power-suppressed ``soft 
symmetry-preserving form factor'' $\xi$ introduced in~\cite{Beneke:2011nf} --- 
performed here within the dispersive sum-rule approach --- considerably 
improves the prediction relative to the agnostic parameterization 
of~\cite{Beneke:2011nf} and the leading-order calculation of   $\xi$ 
in \cite{Braun:2012kp}. Second, the dependence of the form factors
on the shape of the $B$-meson LCDA (which is mostly a dependence on 
$\widehat\sigma_1$ ) is as strong as on $\lambda_B$. Thus any future 
comparison with experiment should aim at the extraction of correlated 
values for $\lambda_B$ and  ``shape parameters'' $\widehat\sigma_1$, etc., 
rather than extracting $\lambda_B$ alone and treating the ``shape 
parameters''  as theoretical uncertainty parameters.

\begin{figure}[t]
\begin{center}
\includegraphics[width=0.435\textwidth]{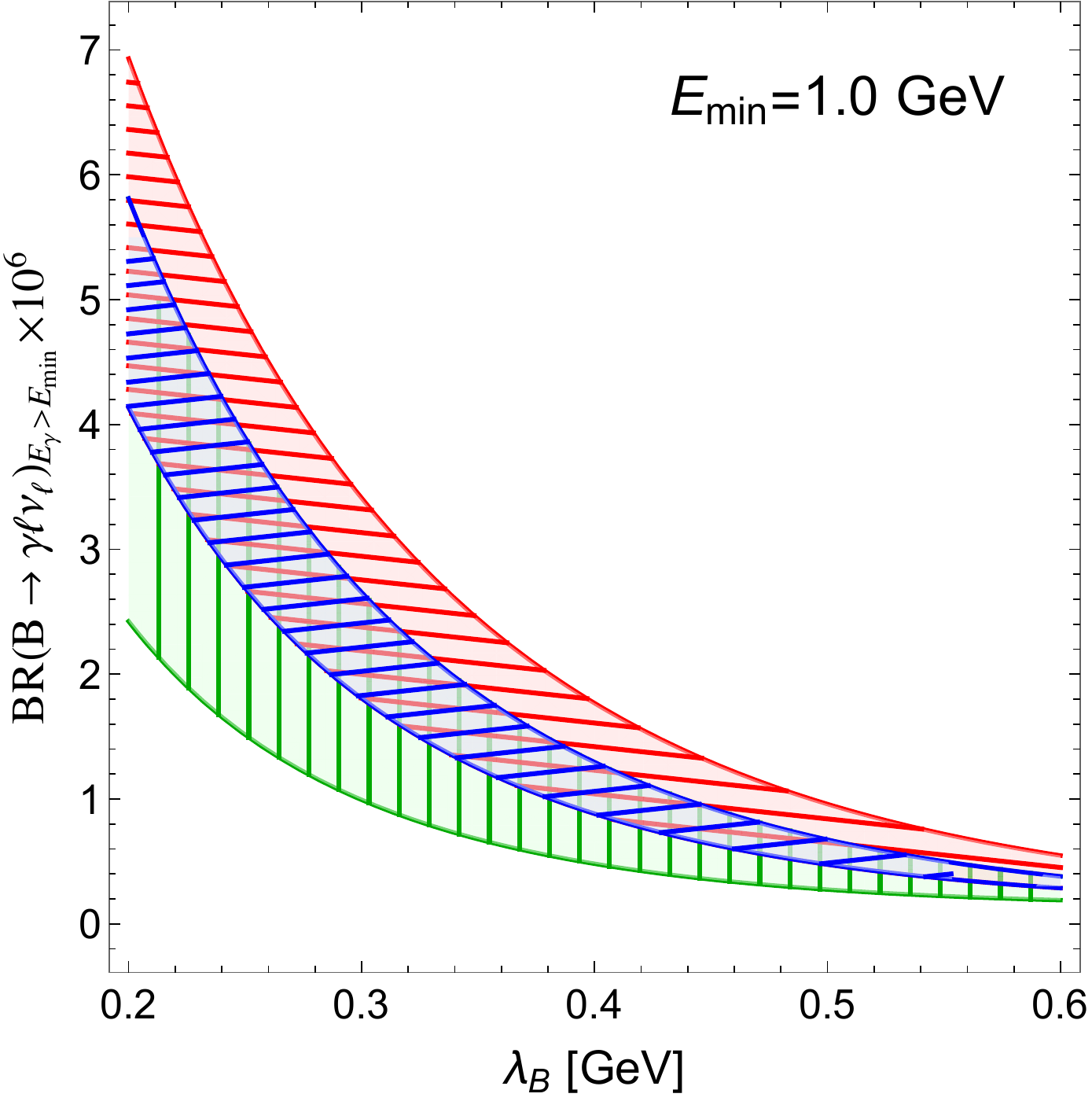}
\hskip0.3cm
\includegraphics[width=0.435\textwidth]{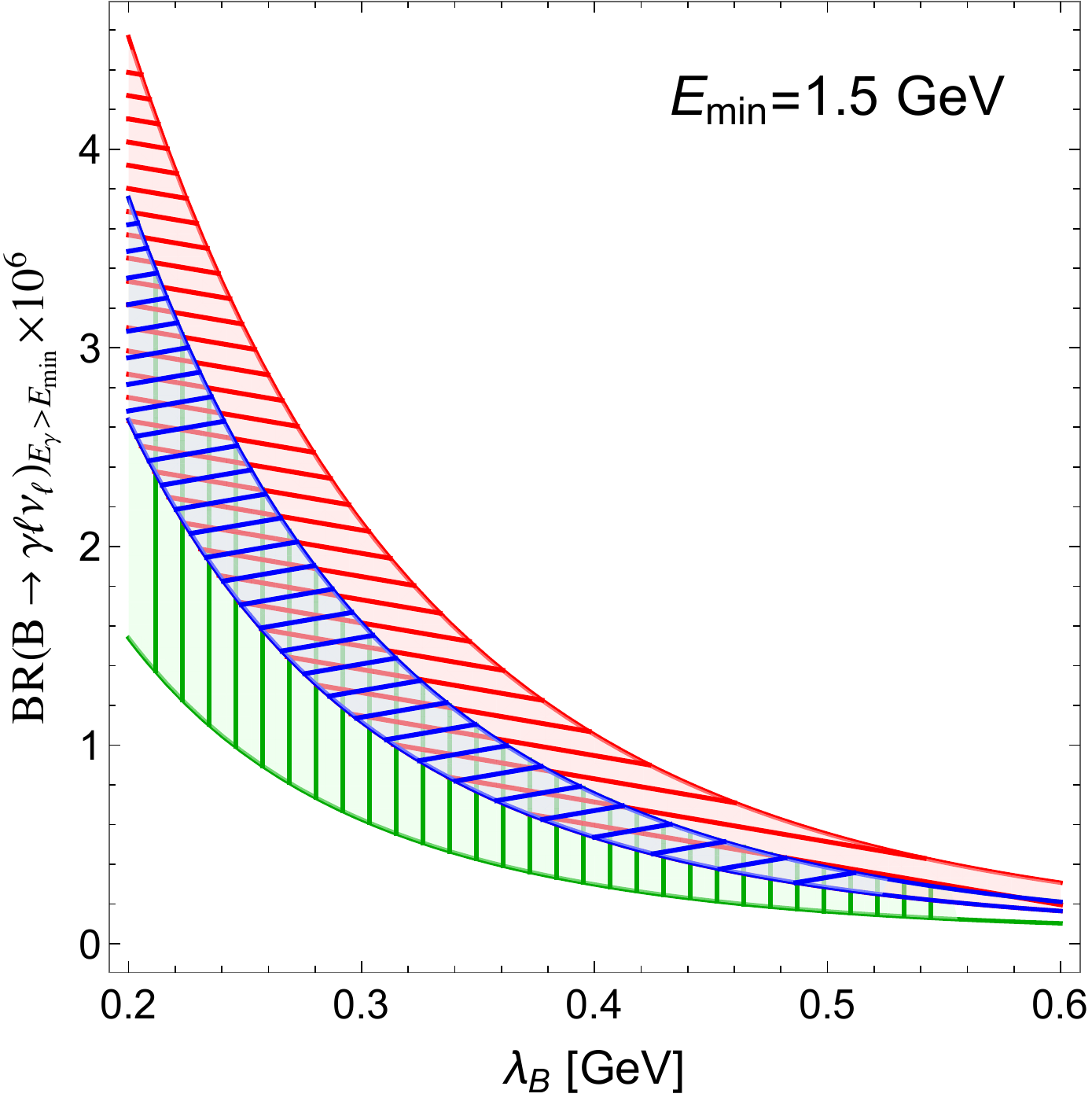}\\[0.3cm]
\includegraphics[width=0.45\textwidth]{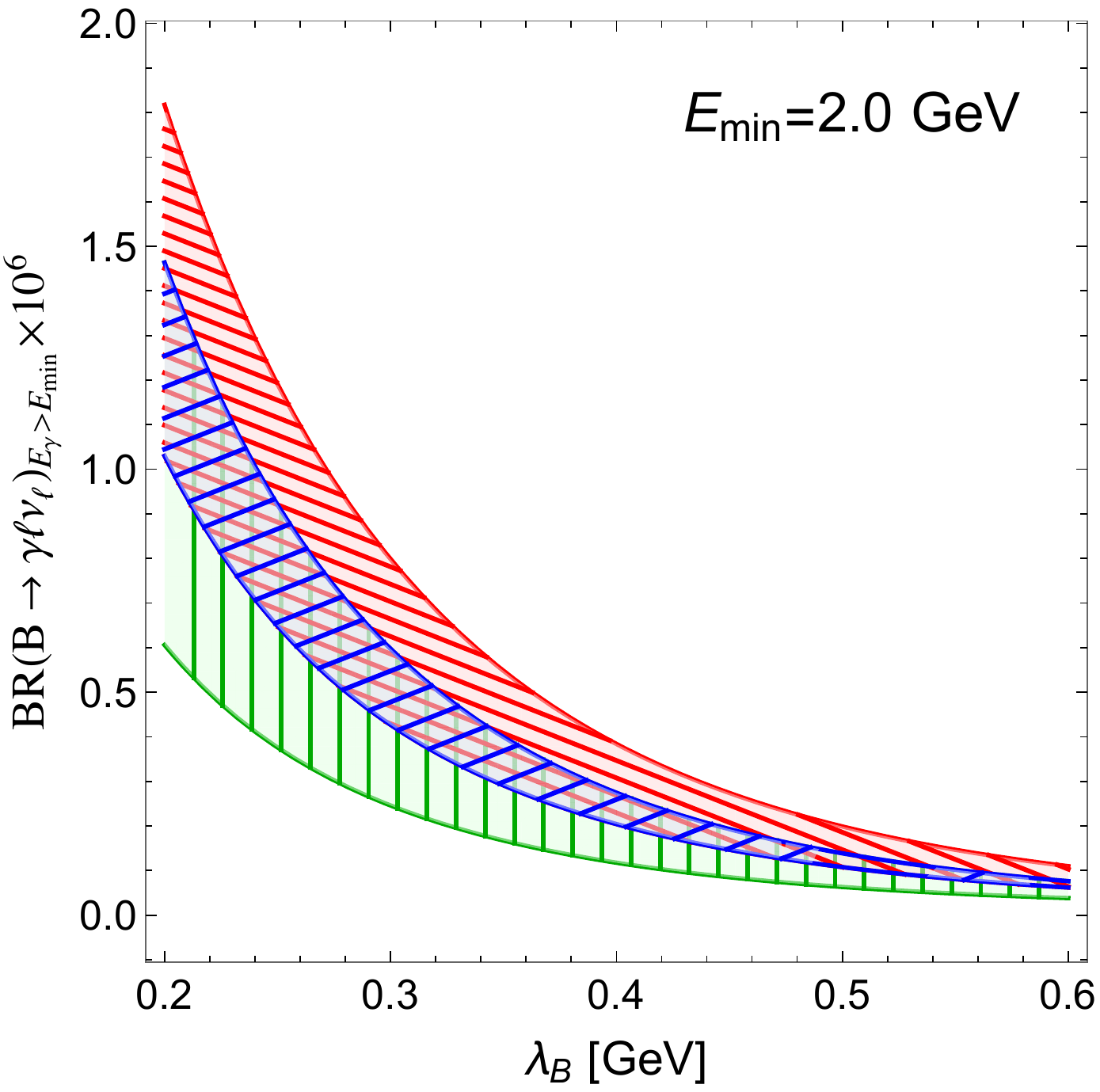}
\caption{\label{fig:rate} 
Integrated partial branching fraction 
$\text{BR}(B \to\gamma\ell\nu_\ell, E_\gamma > E_{\rm min})$ for 
$E_{\rm min} = 1~\text{GeV}$ (middle),  
$E_{\rm min} = 1.5~\text{GeV}$ (middle) and 
$E_{\rm min} = 2~\text{GeV}$ (lower).}
\end{center}
\end{figure}

We finally calculate the partial branching fraction 
$\text{BR}(B \to\gamma\ell\nu_\ell, E_\gamma > E_{\rm min})$, 
integrating \eqref{eq:width} over the photon energy interval 
$E_{\rm min} < E_\gamma < m_B/2$. The result is shown in Fig.~\ref{fig:rate} 
as a function of $\lambda_B$ for three values of the photon energy cut, 
$E_{\rm min} = 1.0~\text {GeV}$, $E_{\rm min} = 1.5~\text {GeV}$ and 
$E_{\rm min} = 2.0~\text {GeV}$ 
with colour coding referring to the three models as discussed above. 
The band corresponds to the variation of $\widehat{\sigma}_1$ such 
that the envelope of all three bands reflects the total  
$\widehat{\sigma}_1$ dependence. For this plot we adopted the 
exclusive $|V_{ub}|$ average, $|V_{ub}| = 
(3.70\pm 0.16)\times10^{-3}$~\cite{Patrignani:2016xqp}, but 
as in the case of $f_B$, we do not include the theoretical uncertainty, 
since the dependence on $f_B |V_{ub}|$ can in principle be eliminated 
by normalizing to another exclusive $b\to u$ decay. The theoretical 
approach requires the photon energy to be large compared to the 
strong interaction scale $\Lambda$. We find that the power corrections 
become increasingly large for smaller $E_\gamma$ such that the expansion 
cannot be considered reliable below $E_\gamma \sim 1.5\,$GeV. Given 
that the first data is statistics-limited \cite{Heller:2015vvm}, it is 
nevertheless tempting to extrapolate to $E_{\rm min} = 1.0~\text {GeV}$, 
and we have done so in Fig.~\ref{fig:rate} --- adding that any 
conclusions drawn from this plot may at best be indicative.

\section{Summary}
\label{sec:summary}

In anticipation of the forthcoming high-statistics measurements of the 
radiative decay  $B\to \gamma \ell \nu_\ell$ by the BELLE II experiment at 
KEK we reconsider its QCD calculation. The interest in this 
decay is mainly due to its distinguished role as the simplest process that 
probes the light-cone $B$-meson distribution amplitude, which in turn is 
an important nonperturbative input in QCD factorization 
for exclusive processes involving $B$ mesons~\cite{Beneke:1999br}. 

The main theoretical issue is to quantify the leading power-suppressed 
effects in $1/E_\gamma$, $1/m_b$, as the leading-power calculation is 
well understood~\cite{Beneke:2011nf}. Following the technique used already 
in~\cite{Braun:2012kp,Wang:2016qii} we employ dispersion relations and 
duality to calculate the power-suppressed soft contributions. In this approach 
soft corrections arise from the modification of the spectral functions 
of the hard-collinear perturbative contributions in the soft region, guided 
by the requirement of a mass gap in the hadronic spectrum
in the photon channel. A strong feature of this technique is that the result 
is insensitive to redefinition of the hard-collinear contributions (e.g. by 
introducing an explicit cutoff for the soft region), which only affects the
decomposition of the answer in hard-collinear and soft contributions but 
leaves their sum intact. 

The present work goes beyond previous 
ones~\cite{Beneke:2011nf,Braun:2012kp,Wang:2016qii} in several directions.
On the technical side, first, we present a calculation of power-suppressed 
higher-twist corrections 
to the form factors that are due to higher Fock states in the $B$-meson and 
to the transverse momentum (virtuality) of the light quark in the valence 
state. These two effects are related by the equations of motion and in the 
sum a rather compact expression can be found, which further reduces to a few constants under rather general assumptions on the form of the 
higher twist LCDAs. The resulting correction to the form factors is negative and not very large, of the order of 10-30\%
depending on the size of the leading contribution. Second, we calculate 
the soft corrections due to twist-five and -six $B$-meson LCDAs 
in the factorization approximation. These terms turn out to be much smaller 
than soft corrections originating from the lower-twist LCDAs, which is 
encouraging, since it indicates that the twist expansion for soft corrections 
is converging.  

On the analysis side, aiming to set the stage for the data analysis once more  
experimental results become available, we present a detailed numerical 
study of the predictions using a rather general class of models for the 
leading-twist LCDA, and the corresponding error analysis. We find that the 
model dependence can be parameterized to a large extent by $\lambda_B$ 
and by the value of the first logarithmic moment $\widehat\sigma_1$ 
(which we redefine, see \eqref{sigma-n}, compared to previous 
studies~\cite{Braun:2003wx,Beneke:2011nf} in order to decorrelate it from 
$\lambda_B$). For a given LCDA, the uncertainty of the calculation of the 
form factors is small, but the dependence of the results on 
$\widehat\sigma_1$ (in addition to the expected dependence on $\lambda_B$) 
is significant. Unless the model space can be constrained otherwise, 
future data should be analyzed in terms of both parameters,
$\lambda_B$ and  $\widehat\sigma_1$.

The most important remaining theory issue in describing 
$B\to \gamma \ell \nu_\ell$ decay for large photon energy is the consistent 
implementation of some version of a cutoff scheme~\cite{Lee:2005gza} together 
with the rederivation of the equation-of-motion relations 
between two-particle and 
three-particle LCDAs in the same framework. This would allow the calculation 
of power-suppressed effects with well-defined moments in the 
presence of a radiatively generated tail of $\phi_+(\omega)$ and, hence, 
to get rid of $\bar\Lambda$ and most of the higher-twist matrix elements as
independent parameters. The two-loop evolution equation for the 
leading-twist $B$-meson LCDA would also be useful for theoretical 
consistency to match the NLL accuracy of the hard and hard-collinear 
evolution.

When this paper was being finalized, Ref.~\cite{Wang:2018wfj} appeared 
suggesting a ``hybrid'' approach where the calculation of the leading-power 
contribution using QCD factorization \cite{Beneke:2011nf} is complemented 
by the calculation of the power-suppressed correction due to photon emission 
from large distances in terms of photon (rather than $B$-meson) LCDAs in the 
LCSR framework. The soft form factor $\xi$ is then entirely independent of 
the parameters of the $B$-meson LCDAs. A potential problem of such ``hybrid'' 
approaches is that the result is not insensitive to the redefinition of 
the perturbative, hard-collinear contributions at sub-leading power accuracy 
(see above). Nevertheless, the validity of this technique and its relation 
to the approach used in the present work are interesting topics for further 
study. The $\gamma^*\gamma\pi$ form factor offers itself as a somewhat 
simpler process where such connections can be investigated.    

\section*{Acknowledgements}
This work  was supported by the DFG grant BR 2021/7-1 
and by the DFG Son\-der\-for\-schungs\-bereich/Trans\-regio~110 
``Symmetries and the Emergence of Structure in QCD''. 
MB wishes to thank the Albert Einstein Center at the University of 
Bern for hospitality when this work was finalized.


\appendix
\section*{Appendix}
\addcontentsline{toc}{section}{Appendices}

\renewcommand{\theequation}{\Alph{section}.\arabic{equation}}
\renewcommand{\thetable}{\Alph{table}}
\setcounter{section}{0}
\setcounter{table}{0}

%
\section{B-meson distribution amplitudes}
\label{App:DAs}
%

Following~\cite{Grozin:1996pq} we define the $B$-meson LCDAs
as matrix elements of the renormalized nonlocal operators built of an 
effective heavy quark field $h_v(0)$ and
a light antiquark at a light-like separation,
\begin{align}
 \langle 0| \bar q(nz) \Gamma [nz,0] h_v(0) |\bar B(v)\rangle &=
-\frac{i}2  F_B\Tr\Big\{\gamma_5 \Gamma P_+ \Big[ \Phi_+(z,\mu)
-\frac{\slashed{n}}{2} \Big(\Phi_+(z,\mu)-\Phi_-(z,\mu)\Big)\Big]
\Big\},
\label{def:two}
\end{align}
where
\begin{equation}\label{WL}
{}[zn,0] \equiv {\rm Pexp}\left[ig\int_0^1\!du\,n_\mu A^\mu(uzn)\right]
\end{equation}
is the Wilson line factor that ensures gauge invariance. Such factors are 
always implied.

Here and below  $v_\mu$ is the heavy quark velocity, $n_\mu$ is a light-like 
vector, $n^2=0$, such that $n\cdot v=1$, $P_+=\frac12(1+\slashed{v})$, 
$\Gamma$ stands for an arbitrary Dirac structure, $|\bar B(v)\rangle$ is 
the $\bar B$-meson state in the heavy quark effective theory (HQET) and 
$F_B(\mu)$ is the scale-dependent HQET decay constant which is related to 
the physical $B$-meson decay constant,
to one-loop accuracy, as
\begin{equation}
\label{defF}
f_B\sqrt{m_B}  = F_B(\mu)K(\mu)  
= F_B(\mu)\left[1+\frac{C_F\alpha_s}{4\pi}\!
\left(3\ln\frac{m_b}{\mu}-2\right)+\ldots\right].
\end{equation}
The parameter $z$ specifies the light antiquark position on the light cone.
To fix the normalization, we assume
\begin{align}
   n_\mu = (1,0,0,1)\,, && \bar n_\mu = (1,0,0-1)\,, 
&& v_\mu = \frac12 (n_\mu + \bar n_\mu)\,, 
&& n\cdot \bar n =2\,.
\label{nbarn}
\end{align}
The functions $\Phi_+(z,\mu)$ and $\Phi_-(z,\mu)$ are the leading- and 
subleading-twist two-particle $B$-meson LCDAs~\cite{Beneke:2000wa}. They are 
analytic functions of $z$ in the lower half-plane $\text{Im}(z)<0$, and are 
related by Fourier transformation to the momentum-space LCDAs
\begin{align}
\Phi_\pm(z,\mu) &= \int_0^\infty d\omega \, 
e^{-i\omega z}\,\phi_\pm(\omega,\mu)\,.
\end{align}
We use upper (lower) case letters for the coordinate-space 
(momentum-space) distributions.

The coordinate-space LCDAs $\Phi_\pm$ can be written in the 
form~\cite{Bell:2013tfa,Braun:2014owa,Braun:2015pha}\footnote{In notation 
of Ref.~\cite{Bell:2013tfa} $s\eta_+(s,\mu) \equiv \rho_+(1/s,\mu)$.} 
\begin{align}
\Phi_+(z,\mu) &= -\frac{1}{z^2}\int_0^\infty ds\, s\,e^{is/z}\, 
\eta_+(s,\mu)\,,
\notag\\[0.3cm]
\Phi_-(z,\mu) &= -\frac{i}{z}\int_0^\infty ds\,  e^{is/z}\, 
\big[\eta_+(s,\mu)+ \eta_3^{(0)}(s,\mu)\big]
= \Phi^{\rm\scriptscriptstyle WW}_-(z,\mu) + \Phi^{\rm t3}_-(z,\mu),
\label{Phi+Phi-}
\end{align}
where $\eta_+(s,\mu)$ and $\eta_3^{(0)}(s,\mu)$ are twist-two and twist-three 
nonperturbative functions that have autonomous scale dependence:
\begin{align}
   \eta_+(s,\mu) &= U_+(s;\mu,\mu_0)\eta_+(s,\mu_0) \,,
\notag\\
   \eta_3^{(0)}(s,\mu) &= r^{N_c/\beta_0}U_+(s;\mu,\mu_0)
   \eta_3^{(0)}(s,\mu_0)\,.
\label{scale-dependence}
\end{align}
Here $r = {\alpha_s(\mu)}/{\alpha_s(\mu_0)}$ and
\begin{eqnarray}
U_+(s;\mu,\mu_0) &=&
\exp\,\biggl\{-\frac{\Gamma_0}{4\beta_0^2}\bigg(
\frac{4\pi}{\alpha_s(\mu_0)}\left[\ln r-1+\frac1r\right]
\nonumber\\[0.2cm]
&& \hspace*{-1.5cm}
-\frac{\beta_1}{2\beta_0}\ln^2r+\left(\frac{\Gamma_1}{\Gamma_0}-\frac{\beta_1}{\beta_0}\right)[r-1-\ln r]\bigg)\biggr\}
\,\left(s\e^{2 \gamma _E}\mu_0\right)^{\frac{\Gamma_0}{2\beta_0}\ln r}\, r^{\frac{\gamma_0}{2\beta_0}}\,,\qquad
\end{eqnarray}
where
\begin{align}
\gamma_0&=-2C_F\, , \qquad
\Gamma_0=4C_F\, ,\qquad\Gamma_1=C_F\Big[\frac{268}3-4\pi^2-\frac{40}9n_f\Big].
\end{align}
The difference with the corresponding expression 
in \cite{Braun:2014owa,Braun:2015pha,Braun:2017liq} is that we included 
the terms in $\beta_1$ and the two-loop cusp anomalous dimension 
$\Gamma_{1}$, which is consistent with resummation to the leading-logarithmic 
accuracy. We further replaced $\mu\to\mu e^{\gamma_E}$ to pass from the 
coordinate-space version of the minimal subtraction scheme used there to 
conventional $\overline{\text{MS}}$ scheme, cf.~\cite{Kawamura:2010tj}.   

In momentum space the exponential factors are substituted by Bessel 
functions, in particular 
\begin{align}
  \phi_+(\omega,\mu) = \int_0^\infty ds\, \sqrt{\omega s}\, J_1(2\sqrt{\omega s}) \eta_+(s,\mu)\,. 
\end{align}
This relation can be inverted to express $\eta_+(s,\mu)$ in terms of 
$\phi_+(\omega,\mu)$:
\begin{align}
   \eta_+(s,\mu) = \int_0^\infty d\omega \, J_1(2\sqrt{s\omega}) \frac{1}{\sqrt{\omega s}}\phi_+(\omega,\mu)\,. 
\end{align}
For the generic ansatz \eqref{Gmodel}, an analytic expression for the LCDA 
$\phi_+(\omega,\mu)$ at arbitrary scale can be found in terms of 
hypergeometric functions using 
\begin{eqnarray}
\lefteqn{
\hspace*{-0.8cm}
 \omega_0 \int_0^\infty ds\, (\omega_0 s)^p \sqrt{\omega s} J_1(2\sqrt{\omega s})\, {}_1F_1(\alpha,\beta,-\omega_0 s) =}
\notag\\&=&
\frac{\omega}{\omega_0} \frac{\Gamma(\beta)\Gamma(2+p) \Gamma(\alpha\!-\!p\!-\!2)}{\Gamma(\alpha)\Gamma(\beta\!-\!p\!-\!2)}
\, {}_2F_2(p+2,p+3-\beta; 2, p+3-\alpha,-\omega/\omega_0)
\notag\\&+&
 \Big(\frac{\omega}{\omega_0}\Big)^{\alpha-p-1}
\frac{\Gamma(\beta)\Gamma(p\!+\!2\!-\!\alpha)}{\Gamma(\beta\!-\!\alpha)\Gamma(\alpha\!-\!p)}
\, {}_2F_2(\alpha,\alpha-\beta+1; \alpha-p-1,\alpha-p,-\omega/\omega_0)\,.
\quad
\end{eqnarray}

Note that the LCDA $\Phi_-(z,\mu)$ is written as a sum of two terms. The 
first one, $\Phi^{\rm\scriptscriptstyle WW}_-(z,\mu)$, is related to the 
leading-twist LCDA $\Phi_+$~\cite{Beneke:2000wa} and is traditionally 
referred to as the Wandzura-Wilczek (WW) contribution. In momentum space 
\begin{align}
  \phi^{\rm\scriptscriptstyle WW}_-(\omega,\mu) = \int_\omega^\infty 
\frac{d\omega'}{\omega'} \phi_+(\omega',\mu)\,.
\label{eq:WWmom}
\end{align}
The second term, $\Phi^{\rm t3}_-(z,\mu)$,  is ``genuinely'' twist-three and 
can be expressed in terms of the three-particle LCDA $\Phi_3$ discussed 
below.   

The three-particle quark-gluon matrix element is parametrized by eight 
invariant functions that can be defined as~\cite{Braun:2017liq}
\begin{eqnarray}
\lefteqn{\langle 0| \bar q(nz_1) g_sG_{\mu\nu}(nz_2)\Gamma h_v(0) 
|\bar B(v)\rangle =}
\nonumber\\[0.15cm]
&=&
\frac12 F_B(\mu) \,\Tr\biggl\{\gamma_5 \Gamma P_+
\biggl[ (v_\mu\gamma_\nu-v_\nu\gamma_\mu)  
\big[{\Psi}_A-{\Psi}_V \big]-i\sigma_{\mu\nu}{\Psi}_V
- (n_\mu v_\nu-n_\nu v_\mu){X}_A
\nonumber\\
&& + \,(n_\mu \gamma_\nu-n_\nu \gamma_\mu)\big[W+{Y}_A\big]
- i\epsilon_{\mu\nu\alpha\beta} n^\alpha v^\beta \gamma_5 \widetilde{X}_A
+ i\epsilon_{\mu\nu\alpha\beta} n^\alpha \gamma^\beta\gamma_5 \widetilde{Y}_A
\nonumber\\[0.1cm]
&& - \,(n_\mu v_\nu-n_\nu v_\mu)\slashed{n}\,{W} + 
(n_\mu \gamma_\nu-n_\nu \gamma_\mu)\slashed{n}\,{Z}
\biggr]\biggr\}(z_1,z_2;\mu)\,.
\label{def:three}
\end{eqnarray}
We use the standard $(+,-,-,-)$ convention for the metric and 
$\gamma_5 = i \gamma^0\gamma^1\gamma^2\gamma^3$. The totally antisymmetric 
Levi-Civita tensor $\epsilon_{\alpha\beta\mu\nu}$ is 
defined with $\epsilon_{0123} =1$. The covariant derivative is defined as
$D_\mu = \partial_\mu -igA_\mu$ and the dual gluon strength tensor as
$\widetilde{G}_{\mu\nu} = \frac12 \epsilon_{\mu\nu\alpha\beta} 
G^{\alpha\beta}$.
The momentum space distributions are defined through 
\begin{align}
 {\Psi}_A (z_1,z_2) &=
\int_0^\infty \!d\omega_1\!  \int_0^\infty \!\!d\omega_2\, 
e^{-i\omega_1 z_1-i\omega_2 z_2}\, {\psi}_A (\omega_1,\omega_2)
\label{3pt-momspace}
\end{align}
and similarly for the other LCDAs.

The invariant functions appearing in the Lorentz structure 
decomposition~\eqref{def:three} can be expanded in contributions of different 
collinear twist. One finds one LCDA of twist three
\begin{align}
\Phi_3 &= {\Psi}_A-{\Psi}_V\,,
\end{align}
and three twist-four LCDAs~\cite{Braun:2017liq}
\begin{align}
  \Phi_4 = {\Psi}_A+{\Psi}_V\,, 
&&
  \Psi_4 = {\Psi}_A+{X}_A\,,
&&
  \widetilde{\Psi}_4 =  {\Psi}_V- \widetilde{X}_A\,.
\end{align}
Neglecting contributions of four-particle operators of the type 
$\bar q GG h_v$ and  $\bar q q\bar q h_v$ the following relation 
holds~\cite{Braun:2017liq}  
\begin{align}
&2 \frac{d}{dz_1}z_1 \Phi_4(z_1,z_2) = \left(\frac{d}{dz_2} z_2+1\right)
\left[\Psi_4 (z_1,z_2)+\widetilde\Psi_4 (z_1,z_2)\right],
\end{align}
or, equivalently,
\begin{align}
&[\psi_4+\widetilde\psi_4](\omega_1,\omega_2)  
- \omega_2\frac{\partial}{\partial \omega_2}
[\psi_4+\widetilde\psi_4](\omega_1,\omega_2) 
= -2  \omega_1\frac{\partial}{\partial \omega_1}\phi_4 (\omega_1,\omega_2)\,,
\label{PhiPsirelation}
\end{align}
so that only two of the three twist-four LCDAs are independent.    

The analysis of the renormalization-group equations for the relevant 
operators~\cite{Braun:2015pha,Braun:2017liq} suggests the following 
representations:  
\begin{eqnarray}
&&\Phi_3(z_1,z_2,\mu) =
\int_0^\infty \!ds \Big[ \eta_3^{(0)}(s,\mu)\,Y_3^{(0)}(s\,|\,z_1,z_2)
+ \frac12 \int_{-\infty}^\infty\!dx\,\eta_3(s,x,\mu)\,
Y_{3}(s,x\,|\,z_1,z_2)\Big],
\notag\\[0.2cm]
 && \Phi_4(z_1,z_2,\mu) =
\frac12 \int^\infty_0{ds}\int^\infty_{-\infty}dx\,
\eta_4^{(+)}(s,x,\mu)\,{Y}_{4;1}^{(+)}(s,x\,|z_1,z_2)\,,
\notag\\[0.2cm]
&& (\Psi_4 \!+\! \widetilde{\Psi}_4)(z_1,z_2,\mu) =
- \int^\infty_0{ds}\int^\infty_{-\infty}dx\,
  \eta_4^{(+)}(s,x,\mu)\,{Y}_{4;2}^{(+)}(s,x\,|z_1,z_2)\,,
\label{Psi+tPsifinal}
\end{eqnarray}
and 
\begin{align}
(\Psi_4 \!-\! \widetilde{\Psi}_4)(z_1,z_2,\mu)&=  
(\Psi_4 \!-\! \widetilde{\Psi}_4)^{\rm t3}(z_1,z_2,\mu)  
+ (\Psi_4 \!-\! \widetilde{\Psi}_4)^{\rm t4}(z_1,z_2,\mu)\,,
\end{align}
where
\begin{eqnarray}
 (\Psi_4 \!-\! \widetilde{\Psi}_4)^{\rm t3} &=&
 2 \int^\infty_0\!{ds}\left(\frac{iz_2}{s}\right)
\biggl[ \eta_3^{(0)}(s,\mu)\,{Y}_{3}^{(0)}(s\,|z_1,z_2) 
\nonumber\\ 
&& +\, \frac12\int^\infty_{-\infty}\!\!dx\,
\eta_3(s,x,\mu)\,{Y}_{3}(s,x\,|z_1,z_2)\biggr]\,,
\notag\\[0.2cm]
(\Psi_4 \!-\! \widetilde{\Psi}_4)^{\rm t4}&=&
 - \int^\infty_0{ds}\int^\infty_{-\infty}dx\, \varkappa_4^{(-)}(s,x,\mu)\, 
{Z}_{4;2}^{(-)}(s,x\,|z_1,z_2)\,.
 \label{Psi-tPsifinal}
\end{eqnarray}
The $Y$- and $Z$-functions in these expressions are eigenfunctions of the 
large-$N_c$ evolutions equations so that the corresponding nonperturbative 
coefficients $\eta_3^{(0)}(s,\mu)$,  $\eta_3(s,x,\mu)$ (twist-three) and 
$\eta_4^{(+)}(s,x,\mu)$, $\varkappa_4^{(-)}(s,x,\mu)$ (twist-four)
have autonomous scale dependence to this accuracy. The function 
$\eta_3^{(0)}(s,\mu)$ is in fact not independent and can be obtained by 
analytic continuation of $\eta_3(s,x,\mu)$ to the complex plane, $x\to i/2$, 
see~\eqref{eta0} below. Explicit expressions for the eigenfunctions in 
coordinate and momentum space, and the corresponding anomalous dimensions 
can be found in~\cite{Braun:2015pha,Braun:2017liq}. Note that the sum 
$(\Psi_4 \!+\! \widetilde{\Psi}_4)$ is purely twist-four, whereas the 
difference $(\Psi_4 \!-\! \widetilde{\Psi}_4)$ contains both the twist-three 
contribution that is related to $\Phi_3$,  and the ``genuine'' twist-four 
part. The two twist-four nonperturbative functions on the line $x=0$ 
are related as
\begin{align}
\left[1 + \partial_s s  - \partial^2_s s^2 - 2 s \bar\Lambda \right]\,   
\eta_+(s,\mu) 
&= \pi  \sqrt{s} \varkappa_4^{(-)}(s,0,\mu) - \pi \sqrt{s}\eta_4^{(+)}
(s,0,\mu)\,.
\label{EOM}
\end{align}
This equation presents the nonlocal generalization of the 
moment relations~\cite{Grozin:1996pq}
\begin{align}
\int_0^\infty\! d\omega\, \omega\,  \phi_+(\omega) = \frac43\bar\Lambda\,, 
\qquad
\int_0^\infty\! d\omega\, \omega^2 \phi_+(\omega) = 2\bar \Lambda^2 
+ \frac23\lambda_E^2 + \frac13 \lambda_H^2\,,
\label{GN}
\end{align}
where $\lambda_E^2$ and $\lambda_H^2$ parametrize the matrix element of 
the local quark-gluon operator
\begin{eqnarray}
\lefteqn{\langle 0| \bar q(0) g_sG_{\mu\nu}(0)\Gamma h_v(0)|\bar B(v)\rangle =}
\nonumber\\&=&
-\frac{i}{6} F_B \lambda^2_H \Tr\Big[\gamma_5\Gamma P_+ \sigma_{\mu\nu}\Big]
-\frac{1}{6} F_B\Big( \lambda^2_H- \lambda^2_E\Big)
  \Tr\Big[\gamma_5\Gamma P_+(v_\mu\gamma_\nu-v_\nu\gamma_\mu)\Big]\,.\quad
\label{def:lambdaEH}
\end{eqnarray}
The matrix element can be estimated from QCD sum rules. One obtains
\begin{align}
\lambda^2_E = 0.11\pm 0.06~\text{GeV}^2, && 
\lambda^2_H =  0.18\pm 0.07~\text{GeV}^2,&& \text{\cite{Grozin:1996pq}}
\\
\lambda^2_E = 0.03\pm 0.02~\text{GeV}^2, && 
\lambda^2_H = 0.06\pm 0.03~\text{GeV}^2, &&  \text{\cite{Nishikawa:2011qk}}
\label{QCDSR:lambdaEH}
\end{align}
where the second calculation takes into account some NLO corrections. Note 
that the ratio
\begin{align}
   R =  \lambda_E^2/ \lambda^2_H \simeq 0.5
\label{REH}
\end{align}
is almost the same in both calculations and is generally more reliable 
than the values of the matrix elements themselves as many uncertainties 
cancel. If the moment relations \eqref{GN} are imposed, 
then, for a given leading twist LCDA $\phi_+(\omega)$, only this ratio 
remains a free parameter.

Physical quantities usually involve an integration over the position of the 
gluon field operator and the resulting expressions 
often become much simpler, e.g.,
\begin{align}
& \int_0^1\! du\ \Big[{\Psi}_4- \widetilde{\Psi}_4\Big]^{\rm t3}(z,uz) 
= z^{-2} \,\Phi^{\rm t3}_-(z)
=  -\frac{i}{z^3}\int_0^\infty ds\,  e^{is/z}\, \eta_3^{(0)}(s,\mu)\,,   
\notag\\[0.2cm]
& \int_0^1d u\, \Big[\Psi_4 - \widetilde{\Psi}_4\Big]^{\rm t4}(z,u z) =
- \frac{i}{z^3}\int^\infty_0{ds}\,\e^{is/z} \, \pi \sqrt{s}  
\varkappa_4^{(-)}(s,0,\mu)\,,   
\label{eq:u-integrals}
\end{align}
and, using \eqref{EOM},
\begin{align}
\int_0^1\!d u\, \Big[\Psi_4 \!-\! \widetilde{\Psi}_4\Big]^{\rm t4}(z,u z) 
&= - \frac{1}{z} \,\biggl[ \,\int_0^1\!du \,u\, \Phi'_+(u z) + \Phi'_+(z) 
+ 2i \bar\Lambda \Phi_+(z)\biggr] 
\notag\\[0.05cm] 
&\quad
-  \int_0^1\!d u\, \bigg\{\Big[\Psi_4\! +\! \widetilde{\Psi}_4\Big](z,u z)
+ \Big[\Psi_4\! +\! \widetilde{\Psi}_4\Big](uz,z)\biggr\}.
\label{EOM2}
\end{align}
A useful representation of the twist-four coefficient functions on the line 
$x=0$ in terms of the momentum-space LCDAs reads
\begin{align}
\eta_4^{(+)}(s,0,\mu) &= \frac{\sqrt{s}}{\pi} 
\int\limits_0^\infty \frac{d\omega_1}{\sqrt{\omega_1}} \int\limits_0^\infty 
\frac{d\omega_2}{\sqrt{\omega_2}} 
\int\limits_0^1\frac{du}{\sqrt{u\bar u}} J_1(2\sqrt{su\omega_1})  
J_1(2\sqrt{s\bar u\omega_2}) [\psi_4+\widetilde\psi_4](\omega_1,\omega_2)\,,
\notag\\
\varkappa_4^{(-)}(s,0,\mu) &= - \frac{\sqrt{s}}{\pi} 
\int\limits_0^\infty \frac{d\omega_1}{\sqrt{\omega_1}} 
\int\limits_0^\infty \frac{d\omega_2}{\sqrt{\omega_2}} 
\int\limits_0^1\frac{u\,du}{\sqrt{u\bar u}} J_1(2\sqrt{su\omega_1})  
J_1(2\sqrt{s\bar u\omega_2}) 
[\psi_4-\widetilde\psi_4]^{\rm tw-4}(\omega_1,\omega_2)\,.
\end{align}

The integrals appearing in the higher-twist corrections 
(\ref{eq:ht1xi}, \ref{eq:ht1dxi}) and (\ref{eq:htmb2}, \ref{eq:htmbd2})  
to  $B\to \gamma \ell \nu_\ell$ can be expressed in terms of the LCDAs 
in the $(s,x)$ representation as follows:
\begin{eqnarray}
&&  \int_0^\infty\!\!{d\omega}\, \ln \omega\, \phi_-^{\rm t3}(\omega) 
= - \int^\infty_0\frac{ds}s\,\eta^{(0)}_3(s,\mu)\,,
\notag\\[0.2cm]
&& \int_0^\infty\!\frac{d\omega_1}{\omega_1 }\! \int_0^\infty 
\frac{d\omega_2}{\omega_1\!+\!\omega_2}
\big[\psi_4\! -\! \widetilde{\psi}_4\big]^{\rm t4}(\omega_1,\omega_2) =
- \pi \int^\infty_0\frac{ds}{\sqrt{s}}\,\varkappa^{(-)}_4(s,0,\,\mu)\,,
\notag\\[0.2cm]
&& \int_0^\infty\!\frac{d\omega_1}{\omega_1 }\! \int_0^\infty \frac{d\omega_2}
{\omega_1\!+\!\omega_2} \phi_3(\omega_1,\omega_2) =
\frac12 \int^\infty_0\frac{ds}s\,\eta_3^{(0)}(s,\mu) 
-\pi\int^\infty_0\frac{ds}s\,\eta_3^{(1)}(s,\mu)\,,\qquad
\label{sx-rep}
\end{eqnarray}
where $\eta_3^{(0)}(s,\mu)$ \eqref{Phi+Phi-} and  $\eta_3^{(1)}(s,\mu)$ are 
the first two coefficients in the Laurent expansion of $\eta_3(s,x,\mu)$ for 
$x\to -i/2$,
\begin{align}
\eta_3(s,x,\mu)\Big|_{x\rightarrow -i/2} 
&= -\frac{i}{\pi}\, \frac{1}{x+i/2} \eta_3^{(0)}(s,\mu) 
+ \eta_3^{(1)}(s,\mu) + \mathcal{O}\Big(x+\frac{i}{2}\Big).  
\label{eta0}
\end{align}    

The leading off-light cone contributions in the current correlation 
functions can be calculated in terms of the 
two-particle higher-twist LCDAs by extending the definition 
from~\cite{Beneke:2000wa} to include ${\cal O}(x^2)$ terms as 
follows:
\begin{eqnarray}
 \langle 0| \bar q(x) \Gamma [x,0] h_v(0) |\bar B(v)\rangle &=&
-\frac{i}2  F_B \Tr\Big[\gamma_5 \Gamma P_+ \Big] 
\int_0^\infty d\omega \, e^{-i\omega (vx)}
\Big\{\phi_+(\omega) + x^2 g_+(\omega)\Big\}
\nonumber\\[0.2cm]
&&{}\hspace*{-4cm} +\frac{i}4  F_B \Tr\Big[\gamma_5 \Gamma P_+ 
\slashed{x} \Big] \frac{1}{vx}
\int_0^\infty d\omega \, e^{-i\omega (vx)} \Big\{[\phi_+-\phi_-](\omega) 
+ x^2 [g_+-g_-](\omega)\Big\}\,.\quad
\label{def:g+g-}
\end{eqnarray}
It is assumed that  $|x^2| \ll 1/\Lambda^2$. The two new LCDAs, $g_+(\omega)$ 
and $g_-(\omega)$ are of twist four and five, respectively. They are not 
independent and can be calculated in terms of the three-particle LCDAs and 
the two-particle LCDAs of lower twist~\cite{Kawamura:2001jm,Braun:2017liq}.
To the tree-level accuracy one obtains in coordinate space
\begin{align}
2 z^2  \mathrm{G}_+(z) & =
-  \Big[ z \frac{d}{dz} - \frac12  + i z \bar \Lambda \Big] \Phi_+(z)
-  \frac{1}{2}\Phi_-(z)
- z^2  \int_0^1\! \bar udu\,{\Psi}_4(z,uz)\,,
\label{KKQT2}
\\[0.1cm]
 2 z^2 \mathrm{G}_-(z)
&= -  \Big[ z \frac{d}{dz} - \frac12  + i z \bar \Lambda \Big] \Phi_-(z) 
- \frac12  \Phi_+(z)
- z^2  \int_0^1\! \bar udu\,{\Psi}_5(z,uz)\,,
\label{KKQT3}
\end{align}
where
\begin{align}
\mathrm{G}_\pm(z,\mu) &= \int_0^\infty d\omega \, 
e^{-i\omega z}g_\pm(\omega,\mu)\,.
\end{align}
In the present context it is important that the expression for the 
two-particle LCDA $G_+(z)$ in~\eqref{KKQT2} and the constraint in~\eqref{EOM} 
are derived under the same assumptions; hence also the relations in 
\eqref{GN} have to be satisfied.    

In practical calculations it proves to be advantageous to write $G_+(z)$ as 
the sum of the Wandzura-Wilczek part and the remaining twist-three and 
twist-four contributions
\begin{align}
\mathrm{G}^{\rm\scriptscriptstyle WW}_+(z) &= - \frac{1}{2 z^2}
\Big[ z \frac{d}{dz} - \frac12  + i z \bar \Lambda \Big] \Phi_+(z) 
- \frac{1}{4 z^2}\Phi^{\rm\scriptscriptstyle WW}_-(z)\,,
\notag\\[0.1cm] 
\mathrm{G}^{\rm t3+t4}_+(z) &= - \frac{1}{4 z^2}\Phi^{\rm t3}_-(z) 
- \frac12 \int_0^1\! \bar udu\,{\Psi}_4(z,uz)\,,
\label{WW}
\end{align}
and to combine the higher-twist terms with the contribution of gluon emission 
from the hard-collinear quark propagator, see~\eqref{eq:Tmn}. In this way, 
remarkable cancellations occur that partially can be expected as a 
consequence of Ward identities.

In~\cite{Braun:2017liq} several models for the higher-twist LCDAs have been 
suggested that incorporate the correct low-momentum behaviour and satisfy the 
(tree-level) EOM constraints. One can show that all these models can be 
obtained as particular cases of a more general ansatz 
\begin{align}
\phi_+(\omega)&= \omega\,f(\omega)\,,
\notag\\
\phi_3(\omega_1,\omega_2)&= -\frac12 \varkappa(\lambda_E^2-\lambda_H^2)\,
\omega_1\omega_2^2\,f'(\omega_1+\omega_2)\,, 
\notag\\
\phi_4(\omega_1,\omega_2)&= \frac12 \varkappa(\lambda_E^2+\lambda_H^2)\,
\omega_2^2\,f(\omega_1+\omega_2)\,,
\label{T4ansatz}
\end{align} 
where $f'(\omega) = df(\omega)/d\omega$ and the normalization constant 
$\varkappa$ is fixed by the leading-twist LCDA through the EOM 
relation~\eqref{GN} to 
\begin{align}
\varkappa^{-1} &= \frac16 \int_0^\infty\!d\omega\,\omega^3 f(\omega) 
= \bar\Lambda^2 + \frac16\big(2\lambda_E^2+\lambda_H^2\big). 
\end{align}
Here it is assumed that the function $f(\omega)$ is normalized as 
$\int_0^\infty d\omega \, \omega\,f(\omega) =1$ and 
decreases sufficiently fast at $\omega\to\infty$ so that 
at least the first three moments $\int_0^\infty d\omega\, 
\omega^{k} f(\omega)$, $k=1,2,3\ldots$ are finite. 
While this cannot hold true in general due to the large-momentum 
tail generated by perturbative radiative corrections~\cite{Lange:2003ff}, 
the assumption is consistent in the context of tree-level calculations 
of higher-twist contributions as performed here. 

From~\eqref{PhiPsirelation} one obtains for this ansatz
\begin{align}
{}[\psi_4+\widetilde\psi_4](\omega_1,\omega_2)&= 
\varkappa(\lambda_E^2+\lambda_H^2)\,\omega_1\omega_2\,f(\omega_1+\omega_2)\,,
\end{align}
but for $[\psi_4-\widetilde\psi_4]$ only the integral \eqref{EOM2} can be determined for a generic profile function $f(\omega)$ 
without additional assumptions. Luckily, only this integral is relevant for $B\to\gamma \ell\nu_\ell$.%
\footnote{For the special choices of the profile function $f(\omega)$ made 
in~\cite{Braun:2017liq} the LCDAs $\psi_4$ and $\widetilde\psi_4$ can be 
separated. One obtains $\psi_4(\omega_1,\omega_2)= 
\varkappa\,\lambda_E^2\,\omega_1\omega_2\,f(\omega_1+\omega_2)$ and
 $\widetilde\psi_4(\omega_1,\omega_2)= \varkappa\,\lambda_H^2\,
\omega_1\omega_2\,f(\omega_1+\omega_2)$.}
The WW part of $\phi_-$ and $g_+$ and the ``genuine'' twist-three 
part of $\phi_-$ can be expressed in terms of $f(\omega)$ in 
the form          
\begin{align}
\phi_-^{\WW}(\omega)&=\int^\infty_\omega\! d\rho\, f(\rho)\,,
\notag\\[0.2cm]
\phi_-^{\rm t3}(\omega)&=\frac16 \varkappa (\lambda_E^2-\lambda_H^2)
\left[\omega^2f'(\omega)+4\omega f(\omega)-2\int^\infty_\omega\! d\rho\, f(\rho)\right]\, ,
\notag\\[0.2cm]
g_+^{\WW}(\omega)&=-\frac14\int^\infty_\omega d\rho\,
\Big(\rho\,\phi_-^{\WW}(\rho)+2(\bar\Lambda-\rho)\phi_+(\rho)\Big)
\notag\\[0.05cm]
&=\frac18\int^\infty_\omega\! d\rho\,(\omega^2+3\rho^2-4\bar\Lambda\rho)
\,f(\rho)\, 
\label{eq:wwpartsinf}
\end{align}
For the particular combinations of the integrated LCDAs entering the 
higher-twist corrections in (\ref{eq:ht1xi}, \ref{eq:ht1dxi}) and 
(\ref{eq:htmb2}, \ref{eq:htmbd2}) we find the remarkably simple results  
\begin{align}
\int_0^\infty\!\!\!\!{d\omega}\, \ln \omega\, \phi_-^{\rm t3}(\omega) &=
\frac16\varkappa
(\lambda_E^2-\lambda_H^2)\, ,
\notag\\[0.2cm]
\int_0^\infty \!\frac{d\omega_1}{\omega_1}  \int_0^\infty \!
\frac{d\omega_2}{\omega_1+\omega_2}\,{\phi}_3(\omega_1,\omega_2)
&=\frac13\varkappa (\lambda_E^2-\lambda_H^2)\,,
\notag\\[0.2cm]
\int_0^\infty\!\frac{d\omega_1}{\omega_1 }\! \int_0^\infty\!\frac{d\omega_2}{\omega_2}
\big[\psi_4\! +\! \widetilde{\psi}_4\big](\omega_1,\omega_2)
& = 2 \int_0^\infty\!\frac{d\omega_2}{\omega_2}\phi_4(0,\omega_2) 
=\varkappa (\lambda_E^2+\lambda_H^2)\,,
\end{align}
which do not depend on the shape of the function $f(\omega)$.
Also the auxiliary functions $\Xi_{1,2}(\omega)$ in \eqref{Xi1}, \eqref{Xi2} 
can be expressed simply as
\begin{align}
\Xi_1(\omega)& = \frac23 \varkappa (\lambda_E^2+2\lambda_H^2)
\left[\omega^2 f(\omega)-2\omega\phi_-^{\WW}(\omega)\right]
-2\omega\phi_-^{\WW}(\omega)+3\omega^2f(\omega)+\omega^3f'(\omega)\,,
\\[-0.1cm]
\Xi_2(\omega)&= - \frac23 \varkappa (\lambda_E^2-\lambda_H^2) 
\left[\omega^2 f(\omega)-2\omega \phi_-^{\WW}(\omega)\right]+ 
(\bar\Lambda-\omega)\,\omega f(\omega)-\omega\phi_-^{\WW}(\omega)
\end{align}
with $\phi_-^{\WW}(\omega)$ given by (\ref{eq:wwpartsinf}) above. 


\end{document}